\documentclass[12pt]{article}
\usepackage{epsfig,amsfonts,amssymb}
\usepackage{hyperref}

\usepackage{cite}
\usepackage{cite}

\usepackage{comment}

\def\ZZZ{{\hbox{ Z\kern-1.6mm Z}}}
\def\RRR{{\hbox{ R\kern-2.4mm R}}}
\def\CCC{{\hbox{ C\kern-2.0mm C}}}
\def\zzz{{\hbox{z\kern-1mm z}}}

\newcommand{\vt}{\vartheta}

\newcommand{\qeq}{{\hbox{=\kern-2.3mm ? \kern.5mm }}}
\renewcommand{\qeq}{=}

\newcommand{\eps}{\epsilon}

\newcommand{\AAA}{{\cal A}}

\newcommand{\OO}{{\cal O}}

\newcommand{\XX}{{\cal X}}
\newcommand{\YY}{{\cal Y}}

\newcommand{\wt}{\widetilde}
\newcommand{\wh}{\widehat}

\newcommand{\NN}{{\cal N}}
\newcommand{\TT}{{\cal T}}

\newcommand{\be}{\begin{equation}}
\newcommand{\ee}{\end{equation}}
\newcommand{\ben}{\begin{eqnarray}\displaystyle}
\newcommand{\een}{\end{eqnarray}}

\newcommand{\refb}[1]{(\ref{#1})}
\newcommand{\p}{\partial}
\newcommand{\sectiono}[1]{\section{#1}\setcounter{equation}{0}}

\def\one{{\hbox{ 1\kern-.8mm l}}}
\def\zero{{\hbox{ 0\kern-1.5mm 0}}}

\newcommand{\bea}[1]{\begin{eqnarray}\label{#1} }
\newcommand{\eea}{\end{eqnarray}}

\newcommand{\eqref}{\refb}

\usepackage{bm}
\usepackage[table]{xcolor}

\begin{document}

\baselineskip 24pt

\begin{center}

{\Large \bf Normalization of Type IIB D-instanton Amplitudes}


\end{center}

\vskip .6cm
\medskip

\vspace*{4.0ex}

\baselineskip=18pt

\centerline{\large \rm Ashoke Sen}

\vspace*{4.0ex}

\centerline{\large \it Harish-Chandra Research Institute, HBNI}
\centerline{\large \it  Chhatnag Road, Jhusi,
Allahabad 211019, India}


\vspace*{1.0ex}
\centerline{\small E-mail:  sen@hri.res.in}

\vspace*{5.0ex}

\centerline{\bf Abstract} \bigskip

We compute the normalization of single D-instanton amplitudes in type
IIB string theory  and show that
the result agrees with the prediction of S-duality due to Green and Gutperle.

\vfill \eject

\tableofcontents

\sectiono{Introduction and summary} \label{s1}

It has been known for many years that string theory amplitudes receive non-perturbative
contribution from D-instantons\cite{9407031,9701093}. 
Many D-instanton induced terms were predicted using S-duality
invariances of various compactified string
theories\cite{9701093,9704145,9706175,9707018,9707241,
9710078,9802090,9808061,9903113,9910055,0411035,
0510027,0708.2950,0712.1252,1001.3647,
1001.2535,1004.0163,1111.2983,1308.1673,1404.2192,1502.03377,1502.03810,1510.07859,
1712.02793}, but except for the early attempts\cite{9701093},
the direct systematic computation of 
these amplitudes from first principles
has not been carried out. However during the last two years progress
was made in the context of two dimensional string 
theory\cite{1907.07688,1912.07170,1908.02782,2003.12076,2012.11624,2101.08566}. 
In particular, \cite{1907.07688,1912.07170} computed the precise
contributions to the amplitudes from D-instantons
in terms of some constants  that
appear to be divergent in the world-sheet formalism. It was then found that 
string field  theory gives finite, 
unambiguous values of
these constants\cite{1908.02782,2003.12076,2012.11624,2101.08566}.

The goal of this paper will be to extend this analysis to type IIB string theory and verify one of the
predictions of S-duality. Our focus will be on the simplest case of ten dimensional type IIB string
theory. Tree level four graviton amplitude in this theory receives a correction proportional  to 
$\zeta(3)$ from an eight derivative
term in the effective action\cite{grosswitten}.  This contribution is not invariant under
S-duality but can be made S-duality invariant by adding one loop and
non-perturbative corrections to the amplitude\cite{9701093,9704145}. The result takes the 
form:\footnote{The prediction of 
S-duality  was shown to be consistent
with some results in $\NN=4$ super Yang-Mills
theory via AdS/CFT correspondence\cite{1912.13365}.  This is also consistent with the analysis
of graviton scattering amplitude using S-matrix bootstrap\cite{2102.02847}.}
\be\label{efirsteq}
{i\over 4} \, 2^6 \, \pi^7 \, g_s^2 \, K_c\,  
\left[2\zeta(3) + {2\pi^2\over 3} \, g_s^2+ 4\, \pi \, g_s^{3/2}\,
\{e^{2\pi i\tau}+e^{-2\pi i\bar\tau}\}
+\cdots
\right]\, \, (2\pi)^{10} \, \delta^{(10)}(k_1+k_2+k_3+k_4)\, ,
\ee
where $\tau=a+i\, g_s^{-1}$, 
$g_s$ is the string coupling defined so that the D-instanton action is given by $2\pi/g_s$,
$a$ is the vacuum expectation value of the RR scalar field
and $K_c$ is a kinematic factor depending on the momenta $\{k_i\}$ and polarizations 
$\{e^{(i)}\}$ of the
external graviton states, as described in \refb{e625}, \refb{e626}. 
The expression \refb{efirsteq} has been written in the string frame, as should be clear from the
explicit factor of $g_s^2$ multiplying the tree level term proportional to $\zeta(3)$. 
In the Einstein frame the expression \refb{efirsteq}
is multiplied by a factor of $g_s^{-7/2}$ and  
becomes proportional to the S-duality invariant function 
$E_{3/2}(\tau,\bar\tau)$\cite{9701093,9704145}.
The one loop term $2\pi^2 g_s^2/3$ in \refb{efirsteq} is known to agree
with the results of explicit
computation\cite{greenschwarz,brink}. In this paper we shall verify 
that the leading non-perturbative term proportional to $e^{2\pi i\tau}$ 
also agrees with the
leading D-instanton contribution to this 
amplitude. 

Formally the leading D-instanton contribution to the four graviton amplitude is given by the product
of four disk amplitudes, each with a single graviton vertex operator and four open
string fermion zero mode insertions\cite{9701093}. 
This part of the amplitude can be computed using straightforward
world-sheet methods. However the amplitude is multiplied by an overall normalization factor
that can be formally identified as the exponential of the annulus amplitude with no vertex
operator insertion. Physically it represents the one loop determinant of the open string fields
on the D-instanton. Due to cancellation between 
the contributions from the NS and R sector states the
annulus partition function vanishes and if we take this literally, it would appear that the
normalization factor is unity. However, this is deceptive since the contribution from the zero modes
cannot be represented as a determinant and the zero mode integrations must be
carried out separately. To deal with this we proceed as follows:
\begin{enumerate}
\item First we show that the exponential of the annulus partition function can be formally 
expressed as an integral over the bosonic and fermionic modes of the open string with
precise normalization. Since there is no subtlety in the non-zero mode sector, the vanishing of
the annulus partition function implies cancellation between the integrals over the non-zero modes
of the open string and we focus on the zero mode sector integrals.
\item Then we show that the integral over the zero modes can be regarded as the result 
of Siegel
gauge
fixing of a gauge invariant integral over the (zero dimensional) 
open string fields. The gauge fixing is done 
following the standard Faddeev-Popov formalism.
\item Some of the zero modes in the gauge fixed version represent bosonic and fermionic
collective modes and must be treated carefully. However one pair of fermionic
zero modes in the NS sector can be identified as the 
Faddeev-Popov ghosts arising from gauge
fixing. The vanishing of the quadratic term of the action of these modes indicate the vanishing of
the Faddeev-Popov determinant and hence the breakdown of the Siegel 
gauge\cite{2002.04043,2006.16270}.
\item We avoid this problem with gauge fixing by using the original gauge invariant version of the
path integral instead of the Siegel gauge fixed version. Since the normalization of the gauge fixed
version was known, this fixes the normalization of the gauge invariant version. This version
does not have integration over the Faddeev-Popov ghost modes, but has an extra integral 
over an out of Siegel gauge mode of the open string. It also has division by the volume of the
gauge group.
\item The out of Siegel gauge mode gives a non-zero contribution to the action. The integration
over this mode takes the form of a Gaussian integral and can be carried out explicitly.
\item We find the volume of the gauge group by relating the string field theory gauge transformation
parameter $\theta$ to the rigid U(1) gauge transformation parameter $\wt\theta$ under which an
open string connecting the original D-instanton to a spectator D-instanton picks up a phase
$e^{i\wt\theta}$. This relationship is found by comparing the gauge transformation laws in string
field theory to the rigid U(1) gauge transformation laws. Once this is done we can express the
integration over $\theta$ in terms of 
integration over $\wt\theta$ and then use the fact that $\wt\theta$ has
period $2\pi$ to compute the volume of the gauge group.
\item The remaining modes in the NS sector represent bosonic zero modes related to collective
modes of the D-instanton describing its location in space-time. We determine the precise 
normalization relating the two sets of 
modes by comparing the coupling of the open string zero modes to
closed strings to the expected coupling of the collective modes to closed strings. 
Using this we can express
the integration over these bosonic zero modes in terms of integration over the collective modes
with some  specific normalization factor. The integration over the collective modes is left aside,
to be done at the end after combining the contribution from all the pieces. The final
integration over these modes generate the
usual energy-momentum conserving delta function $(2\pi)^{10}\, \delta^{(10)}(\sum_ip_i)$.
\item In the R sector there are 16 fermion zero modes, and all of these can be related to the
fermionic collective modes of the D-instanton associated with broken supersymmetry. Integration
over these modes is also set aside till the end after we combine all the pieces. In particular,
we need to insert 16 fermionic modes into the four disks, each carrying a single graviton vertex
operator. The integration over the fermionic collective modes now produces a suitable 16-dimensional
$\epsilon$ tensor that needs to be combined with the rest of the amplitude.
\end{enumerate}

The answer for the one instanton contribution to the 
four graviton amplitude, computed this way, takes the form:
\be \label{efinresult}
 i\, e^{2\pi  ia}\, e^{-2\pi/g_s} \, 2^{6}\,
\pi^{8}\, g_s^{7/2}\, K_c \, (2\pi)^{10} \, \delta^{(10)}(k_1+k_2+k_3+k_4)\, .
\ee
\refb{efinresult} agrees with the term proportional to $e^{2\pi i\tau}$ in 
\refb{efirsteq}.
$a$ dependence of the amplitude can be obtained by exponentiating the disk one point function
of the RR scalar field since that is the only amplitude that involves $a$ and not its derivative. 
The overall phase of the
term is not determined due to the usual ambiguities in evaluating path integral over chiral fermions,
but this phase can be absorbed into a shift of $a$.

The fact that the instanton contribution gives the correct dependence
on $g_s$ was already noted in \cite{bryappear}. 
The ratio of the subleading non-perturbative corrections, hidden in the $\cdots$ in \refb{efirsteq}, 
to the leading 
non-perturbative correction
is also being
analyzed in \cite{bryappear}.

The rest of the paper is organized as follows. In \S\ref{s2} we describe 
our normalization conventions
in the world-sheet string theory and compare them with those of
\cite{polchinski} whose results we use. In \S\ref{s3} we describe our normalization conventions
in string field theory, and compare the coupling constants and fields that arise there with those
appearing in \cite{polchinski}. Sections \ref{s4}-\ref{s6} contain the main results of this paper.
In \S\ref{s4} we compute the normalization of the D-instanton 
amplitudes by manipulating the exponential of the annulus zero point function following the
procedure described earlier in this section. 
This computes the total contribution from the steepest
descent contour passing through 
the instanton. However the actual contribution of the instanton to the full
amplitude depends on how the steepest descent contour fits inside the actual integration
contour. This produces a multiplier factor that accompanies the normalization. In \S\ref{s5}
we argue that for the D-instanton of type IIB string theory this multiplier factor is one. In 
\S\ref{s6} we compute the disk amplitude with one graviton and four fermionic open string
zero mode insertions and combine this with the result of \S\ref{s4} to compute the leading
D-instanton contribution to the four graviton amplitude. In \S\ref{s7} we review the
prediction of S-duality for this amplitude and show that the result of explicit D-instanton
calculation agrees with the prediction of S-duality. In \S\ref{s8} we discuss possible generalization
of this analysis to D-instanton contribution in other (compactified) string theories, including the 
contribution from Euclidean D-branes wrapped along compact cycles.

\sectiono{Conventions for the world-sheet theory} \label{s2}

In this section we shall describe our normalization conventions. Since we are trying to
reproduce a single constant, it is important that we carefully keep track of all the constants in
our analysis. We work in the $\alpha'=1$ unit. 
For the rest of the conventions, we shall try to follow closely the ones used in 
\cite{1703.06410}. In a few places we shall differ from the convention of 
\cite{1703.06410}; we shall
mention them as we encounter these differences.

The world-sheet
of type IIB string theory has a set of 10 scalar fields $X^\mu$ describing the target space-time 
coordinates, their superpartner left and right-moving fermions $\psi^\mu$, $\bar\psi^\mu$, the
world-sheet grassmann odd ghost fields $b$, $c$, $\bar b$, $\bar c$ and the grassmann even
ghost fields $\beta,\gamma,\bar\beta,\bar\gamma$. The $\beta$,$\gamma$ system is `bosonized'
by introducing scalar fields $\phi,\bar\phi$, and fermionic fields $\xi,\eta,\bar\xi,\bar\eta$ via the
relations:
\be\label{ebosonize}
\beta=\p\xi\, e^{-\phi}, \quad \gamma=\eta\, e^\phi, \quad \bar\beta=\bar\p\bar\xi\, e^{-\bar\phi}, 
\quad \bar\gamma=\bar\eta\, e^{\bar\phi}\, .
\ee
The operator products between various fields take the form:
\ben \label{ematterope}
&& \p X^\mu(z) \p X^\nu(w) = -{\eta^{\mu\nu}\over 2 (z-w)^2}+\cdots, \quad
\psi^\mu (z) \psi^\nu(w) = -{\eta^{\mu\nu}\over 2(z-w)}+\cdots\, , \nonumber \\
&& c(z) b(w) =(z-w)^{-1}+\cdots, \nonumber \\
&& \xi(z)\eta(w) = (z-w)^{-1}+\cdots, \nonumber \\
&& e^{q_1\phi(z)} e^{q_2\phi(w)} = (z-w)^{-q_1q_2} e^{(q_1+q_2)\phi(w)}+ \cdots \, , \nonumber \\
&& \p \phi (z)\,  \p\phi(w) = -{1\over (z-w)^2} +\cdots \, ,
\een
where $\cdots$ denote less singular terms whose knowledge will not be
needed for our analysis. The Minkowski metric $\eta^{\mu\nu}$ is taken to have
mostly + signature, and is replaced by $\delta_{\mu\nu}$ in the euclidean computation.
There are similar operator product expansions involving anti-holomorphic
fields that we have not written down. 
In the following discussion we shall only write down the various relations involving the holomorphic
fields, with the implicit understanding that there are similar relations involving anti-holomorphic
fields as well.

We assign ghost number 1 to $c,\bar c,\gamma,\bar\gamma,\eta,\bar\eta$, $-1$ to
$b,\bar b,\beta,\bar\beta,\xi,\bar\xi$ and 0 to the rest of the fields. We also assign
picture number $q$ to $e^{q\phi}$ and $e^{q\bar\phi}$, 1 to $\xi,\bar\xi$,
$-1$ to $\eta,\bar\eta$ and 0 to the rest of the
fields. The SL(2,C) invariant vacuum carries zero ghost number and picture number.

The stress tensor $T(z)$ and its fermionic partner 
$T_F(z)$ for the matter sector take the form:
\be \label{emattertdef}
T_m(z) = - \partial X^\mu \partial X^\nu \eta_{\mu\nu} +
\psi_\mu \p \psi^\mu, \quad
T_F(z) = -\psi_\mu \p X^\mu\, ,
\ee
with similar expressions for their anti-holomorphic counterparts. 
The operator product expansions involving $T_m$ and $T_F$ take the form:
\ben\label{etope}
&& T_m(z) T_m(w) = {15\over 2} {1\over (z-w)^4}+{2\over (z-w)^2} T_m(w) +{1\over z-w}
\p T_m(w)+\cdots, \nonumber\\
&& T_F(z) T_F(w) 
=  {5\over 2} \, {1\over (z-w)^3}+ {1\over 2}\, {1\over z-w} T_m(w) + \cdots, \nonumber \\ 
&&
T_m(z) T_F(w) = {3\over 2} {1\over (z-w)^2} T_F(w) + {1\over z-w} \p T_F(w)
+\cdots\, .
\een
The stress tensors of the ghost fields are given by
\be
T_{b,c}=-2 \, b\, \p\, c + c\, \p\, b, \quad 
T_{\beta,\gamma} (z) = {3\over 2} \beta\p\gamma + {1\over 2} \gamma\p\beta
= T_\phi + T_{\eta,\xi}\, ,
\ee
where
\be \label{e2.4}
T_{\eta,\xi} = -\eta \p \xi\, ,
\qquad T_\phi = -{1\over 2} \p\phi \p\phi - \p^2 \phi\, .
\ee
The BRST charge is given by
\be \label{ebrs1}
Q_B = \ointop dz \jmath_B(z) \, ,
\ee
where
\be \label{ebrstcurrent}
\jmath_B(z) =c(z) \{T_{m}(z) + T_{\beta,\gamma}(z) \}+ \gamma (z) T_F(z) 
+ b(z) c(z) \p c(z) 
-{1\over 4} \gamma(z)^2 b(z)\, ,
\ee
and $\ointop$ is normalized to include the $1/(2\pi i)$ factor so that 
$\ointop dz/z=1$.

The picture changing operator (PCO) $\XX$\cite{FMS,verlinde}
will be taken to be:
\be \label{epicture}
\XX(z) = 2\, \{Q_B, \xi(z)\} =2\,  c \, \partial \xi + 
2\, e^\phi T_F - {1\over 2} \p \eta \, e^{2\phi} \, b
- {1\over 2} \p\left(\eta \, e^{2\phi} \, b\right)\, .
\ee
This differs from the one used in \cite{1703.06410} 
by a factor of 2. Since the picture number
non-conservation on a Riemann surface of genus $g$ 
is proportional to $2g-2$, and since string amplitudes carry factors of $g_s^{2g-2}$ where $g_s$
is the string coupling constant, the difference in the normalization of the PCO can be 
absorbed into a redefinition of the string coupling and the normalization of the vertex operators.
We shall see that \refb{epicture} 
is a convenient normalization to use for computation of amplitudes.

We also introduce the inverse picture changing operator
\be \label{eydef}
\YY = 2\, c\, \p\xi \, e^{-2\phi} \, .
\ee
Both $\XX$ and $\YY$ commute with the BRST operator. Furthermore, they have a
non-singular 
operator product expansion:
\be 
\YY(z) \XX(w) = 1 + \OO(z-w)\, .
\ee

Since we shall be using some of the results from \cite{polchinski} we shall now give the relation
between the normalization conventions used here and those used in \cite{polchinski}. The results
of \cite{polchinski} can be found by making the following replacements in our 
formul\ae:\footnote{With these replacements, the bosonization rule for $\beta,\gamma$
should take the form $\beta=e^{-\phi}\, \p\xi$, $\gamma=\eta\, e^\phi$. 
Ref.\cite{polchinski} states the bosonization rules as
$\beta=e^{-\phi}\, \p\xi$, $\gamma=e^\phi\, \eta$, 
but this is inconsistent
with the operator product expansion $\gamma(z)\beta(w)\simeq (z-w)^{-1}$ used in
\cite{polchinski} if we take $\xi$, $\eta$ to anti-commute with $e^{\pm\phi}$. }
\ben \label{ecompare}
&& \beta\to -\beta/2, \quad \gamma\to 2\, \gamma, \quad \xi\to \xi/2, \quad \eta\to 2\eta, \quad
\phi\to\phi,\nonumber \\
&&X^\mu\to X^\mu, \quad \psi^\mu \to -i\, \psi^\mu/\sqrt 2, \quad T_m\to T_B, \quad T_F\to T_F/2\, .
\een

Next we introduce the 16-component 
spin fields $S^\alpha$ and $S_\alpha$ in the matter sector, carrying opposite
chirality. We shall use the convention that $e^{-\phi/2}S_\alpha$ and $e^{-3\phi/2}S^\beta$ are
GSO even operators. The relevant operator product involving the spin fields are:
\ben\label{espinope}
\psi^\mu(z) \ e^{-\phi/2} S_\alpha(w) &=& {i\over 2}\, (z-w)^{-1/2}\, 
(\gamma^\mu)_{\alpha\beta} e^{-\phi/2}\, S^\beta(w) 
+\cdots , \nonumber\\
\psi^\mu(z) \ e^{-\phi/2} S^\alpha(w) &=& {i\over 2}\, (z-w)^{-1/2}\, 
(\gamma^\mu)^{\alpha\beta} e^{-\phi/2}\, S_\beta(w) 
+\cdots \, ,
\nonumber\\
e^{-3\phi/2} S^\alpha(z)\ e^{-\phi/2} S_\beta(w)   &=& (z-w)^{-2} \, \delta^\alpha_\beta \,
e^{-2\phi}(w)+\cdots ,
\nonumber \\
e^{-\phi/2} S_\alpha(z)  \ e^{-\phi/2} S_\beta(w) &=& i\, (z-w)^{-1} \, 
(\gamma^\mu)_{\alpha\beta} \, e^{-\phi} \, \psi_\mu(w) +\cdots \, ,
\een
where the $16\times 16$ matrices $\gamma^\mu_{\alpha\beta}$ satisfy the identities:
\ben
&& (\gamma^i)^{\alpha\beta}=(\gamma^i)^{\beta\alpha}, \quad
(\gamma^i)^{\alpha\beta}=(\gamma^i)_{\alpha\beta}, \quad \{\gamma^i,\gamma^j\}
= 2\, \delta_{ij}, \quad \hbox{for $1\le i\le 9$}\, ,\nonumber \\
&& (\gamma^0)^{\alpha\beta}=\delta_{\alpha\beta}, \qquad (\gamma^0)_{\alpha\beta}=-
\delta_{\alpha\beta}\, .
\een
These are related to the full
ten dimensional gamma matrices $\Gamma^\mu$ as follows:
\be 
\Gamma^\mu = \pmatrix{0 & (\gamma^\mu)^{\alpha\beta}\cr (\gamma^\mu)_{\alpha\beta} & 0}\, .
\ee
An explicit choice of such gamma matrices can be found {\it e.g.} in appendix A of
\cite{0302147}.
It will be understood that when we take product of the $\gamma^\mu$'s, the successive 
$\gamma^\mu$'s will have their indices alternating between upper and lower indices. Therefore
$(\gamma^\mu\gamma^\nu)^\alpha_{~\beta}$ will correspond to $(\gamma^\mu)^{\alpha\delta}
(\gamma^\nu)_{\delta\beta}$. With this convention, we have
\be
\{\gamma^\mu, \gamma^\nu\} = 2\, \eta^{\mu\nu}\, I_{16}\, ,
\ee
where $I_{16}$ denotes the $16\times 16$ identity matrix.
The consistency of \refb{espinope} with \refb{ematterope} can be seen by studying various 
correlation functions. For example, we have
\be
\langle c e^{-\phi} \psi^\mu(z_1) c e^{-\phi/2} S_\alpha(z_2) c e^{-\phi/2} S_\beta(z_3)\rangle
=i\, K\, \gamma^\mu_{\alpha\beta}/2\, ,
\ee
where $K$ is an overall constant giving $\langle c\p c\p^2 c e^{-2\phi} \rangle/2$ 
in the holomorphic sector. This can be obtained by either taking the operator product of
the second and third operators first using \refb{espinope} and then using \refb{ematterope},
or by taking the operator product of the first and the second operator first using
\refb{espinope} and then using \refb{espinope} again.

We now give the mode expansion of the various fields.
The ghost and the matter fields have mode expansions
\ben \label{emodeexpan}
&& b(z) =\sum b_n z^{-n-2}, \quad c(z) = \sum_n c_n z^{-n+1}, \nonumber \\
&& \beta(z) =\sum_n \beta_n z^{-n-{3\over 2}}, \quad \gamma(z) = \sum_n \gamma_n
z^{-n+{1\over 2}},
\quad \eta(z) =\sum_n \eta_n z^{-n-1}, \quad \xi(z) = \sum_n \xi_n z^{-n}\, ,
\nonumber \\
&& i\, \sqrt 2\, \p X^\mu(z) = \sum_n \alpha^\mu_n z^{-n-1}, \qquad
i\, \sqrt 2\, \psi^\mu(z) = \sum_n d^\mu_n z^{-n-1/2}\, .
\een
Also useful will be the mode expansions of the total stress tensors of the matter and
ghost
superconformal field theory and the super-stress tensor of the matter theory:
\be \label{evira}
T(z) =\sum L_n z^{-n-2}\, , \qquad
T_F(z) ={1\over 2}\sum_n G_n^{(m)} \, z^{-n-3/2}\, .
\ee
Note that in this equation $T(z)$ refers to the total stress tensor of all the fields, while
$T_F$ is the super-stress tensor of the matter fields only. The superscript $(m)$
of $G_n^{(m)}$ will serve to remind us of this.

The normalization conventions described above will be used for both closed and open
strings For open strings the expansion coefficients of the anti-holomorphic fields are not independent,
but are
related to those of the holomorphic fields. For computing correlation functions on the
upper half plane, this relationship is used to arrive at the doubling trick in which we replace
the upper half plane by the full complex plane and the anti-holomorphic fields in the
upper half plane by holomorphic fields at the complex conjugate points. 

Finally, we state the normalization of the vacua of the closed string and
the open string. For the closed string vacuum carrying momentum $k$, we
choose the normalization\cite{1703.06410}:
\be\label{eclosednorm}
\langle k| c_{-1}\bar c_{-1} c_0\bar c_0 c_1 \bar c_1\, e^{-2\phi}(0) e^{-2\bar\phi}(0)|k'\rangle=
- (2\pi)^{10}\delta^{(10)}(k+k')\, .
\ee
The normalization of the open string vacuum on a $p$-brane will be chosen as:
\be\label{eopennorm}
\langle k| c_{-1} c_0 c_1 \, e^{-2\phi}(0)|k'\rangle=(2\pi)^{p+1} \delta^{(p+1)}(k+k')\, .
\ee

\sectiono{Conventions for string field theory} \label{s3}

We shall now review some of the relevant properties of open-closed superstring field theory that
describes the coupled dynamics of the degrees of freedom of a D-$p$-brane and the closed
string degrees of freedom. We shall need only a small part of the string field theory and not
the full details. The full details can be found in \cite{9705241,1907.10632}, 
but our convention differs from that of 
\cite{1907.10632} in one important way. In the analysis of 
\cite{1907.10632} the kinetic term of the closed string fields
was accompanied by a factor of $g_s^{-2}$, that of the open string fields was accompanied by
a factor of $g_s^{-1}$ and the normalization of the interaction terms were specified only implicitly
by requiring that they satisfy appropriate sewing identities. Here we shall accompany the kinetic term
of the closed string fields by a constant $\kappa^{-2}$ and that of the open string fields by a 
different constant $g_o^{-1}$ and adjust the relation between $\kappa$ and $g_o$ so that the
interaction terms have simple normalization. This corresponds to appropriate 
rescaling of the closed and the
open string fields. We shall introduce a third constant $g_s$ such that the tension of a BPS
D-$p$-brane is given by $(2\pi)^{-p}/g_s$. In particular the type IIB D-instanton action will be
given by $2\pi/g_s$.

\subsection{Closed string sector of string field theory} \label{s2.2}

We shall begin by writing down the kinetic term and the sphere 3-point interaction terms for the
NSNS sector classical closed string field. 
We denote the NSNS sector classical
closed string field by a state $|\psi_c\rangle$ in the NSNS sector of the 
closed string Hilbert space of ghost number 2, satisfying,
\be
(b_0-\bar b_0)|\psi_c\rangle=0, \qquad (L_0-\bar L_0)|\psi_c\rangle=0\, ,
\ee
and write the quadratic and the cubic term in the action as:
\be \label{eclosedaction}
S_c={4\over \kappa^2} \left( {1\over 2} \langle\psi_c|c_0^- (Q_B+\overline Q_B)|\psi_c\rangle + {1\over 3!}
\{\psi_c^3\}\right), \quad c_0^-\equiv(c_0-\bar c_0)/2\, ,
\ee
where $\{V_1V_2V_3\}$ is given by
the sphere correlation function of a pair of PCOs and three closed string
vertex operators $V_1$, $V_2$, $V_3$, 
inserted using appropriate local coordinate system specified by string field
theory. The correlation function is
computed with the
normalization \refb{eclosednorm}. 
Our sign convention for the action is such that in the
Euclidean (Lorentzian) theory we take the weight factor in the path integral to be $e^{S}$
($e^{iS}$). 
The string field $|\phi_c\rangle$ with canonical normalization is related to
$\psi_c$ via
\be
|\psi_c\rangle =\kappa|\phi_c\rangle\, , 
\ee
so that 
\be \label{e3.4}
S_c={4} \left( {1\over 2} \langle\phi_c|c_0^- (Q_B+\overline Q_B)|\phi_c\rangle + {\kappa\over 3!}
\{\phi_c^3\}\right), \quad c_0^-\equiv(c_0-\bar c_0)/2\, ,
\ee
To check that the kinetic term has the correct normalization, we can fix Siegel gauge 
$b_0|\phi_c\rangle=0$. In this gauge we can replace $Q_B+\overline Q_B$ 
by $c_0L_0+\bar c_0\bar L_0$,
and the kinetic term of the action reduces to:
\be\label{e130}
\langle\phi_c|c_0\bar c_0 (L_0 +\bar L_0)|\phi_c\rangle\, .
\ee
Since each of $L_0$ and $\bar L_0$ have additive terms $k^2/4$, the kinetic term has 
the correct normalization $k^2/2$. In particular, if we define the graviton field
$h_{\mu\nu}$ as the following term in the expansion of $|\phi_c\rangle$:
\be \label{efieldexp}
-\int {d^{10}k\over (2\pi)^{10}} \, h_{\mu\nu}(k) \, c_1\bar c_1 d^\mu_{-1}\bar d^\nu_{-1} \, 
e^{-\phi}(0) e^{-\bar\phi}(0)|k\rangle\, ,
\ee
then, with the normalization \refb{eclosednorm}, 
the kinetic term for $h_{\mu\nu}$ will take the form:
\be
-{1\over 2} \, \int {d^{10}k\over (2\pi)^{10}} \, h_{\mu\nu}(-k)\, k^2 \, h^{\mu\nu}(k)\, .
\ee
This agrees with the quadratic term in the Einstein action,
\be\label{eeff1}
{1\over 2\, \kappa^2}\int d^{10}x\,  \sqrt{-\det g}\, R\, ,
\ee
in the de Donder gauge, if we expand the metric as
\be\label{eeff2}
g_{\mu\nu}=\eta_{\mu\nu} + 2\, \kappa\, \int  {d^{10}k\over (2\pi)^{10}} \, h_{\mu\nu}(k)\, e^{ik.x}\, .
\ee

In
this convention, a normalized graviton state of momentum $k$ and polarization $e_{\mu\nu}$
in the $(-1,-1)$ picture has the form:
\be \label{eedefn}
-e_{\mu\nu} \, c_1\, \bar c_1\, d^\mu_{-1} \bar d^\nu_{-1} \, e^{-\phi}(0)
\, e^{-\bar\phi}(0)  |k\rangle\,, \qquad e_{\mu\nu}=e_{\nu\mu},
\quad \eta^{\mu\nu}e_{\mu\nu}=0, \quad k^\mu\, e_{\mu\nu}=0, \quad 
e^{\mu\nu} e_{\mu\nu}=1\, .
\ee
Using \refb{emodeexpan}, the associated vertex operator is given by
\be \label{egravminone}
V=-2\, e_{\mu\nu} \, c\, \bar c\, e^{-\phi} \, \psi^{\mu} \, e^{-\bar\phi}\, \bar\psi^\nu\, e^{ik.X}\, .
\ee
We shall also need the zero picture vertex operator of this state, obtained by multiplying this by
the picture changing operators $\XX\, \bar\XX$. This takes the form:
\be\label{e125}
2\, e_{\mu\nu} \, \bar c\, c\,  \left\{\p X^{\mu} + i\, k_\rho\, \psi^\rho\psi^\mu\right\} \, 
\left\{\bar\p X^{\nu} + i\, k_\sigma\, \bar\psi^\sigma\bar\psi^\nu\right\} \, e^{ik.X}+\cdots\, ,
\ee
where $\cdots$ involves terms proportional to $\gamma\, \psi^\mu$ and $\bar\gamma\bar\psi^\nu$
that will not be needed for our analysis.
This agrees with the conventions of \cite{polchinski}
after using the translation rules \refb{ecompare}.

We shall now argue that with this normalization the three point functions of the gravitons also 
agree with that of \cite{polchinski}.
For this let us consider three gravitons with momenta $\{k_i\}$ and 
polarizations $e^{(i)}_{\mu\nu}$ for $1\le i\le 3$. Let us denote by $V_i$'s their vertex operators.
Comparing \refb{efieldexp} with \refb{eedefn} we see that the $V_i$'s are
given as in \refb{egravminone}. It now follows from \refb{e3.4}
that the three graviton amplitude is given by
\be\label{egrav}
4\, i\, \kappa\,\{ V_1 V_2 V_3\} = 4\, i\, \kappa\, \langle V_1(z_1) V_2(z_2) V_3(z_3)\rangle \, ,
\ee
where $\langle~\rangle$ denotes correlation function on the sphere
and $z_1$, $z_2$ and $z_3$ are three fixed points on the sphere. The factor of
$i$ is the standard factor that
arises in the computation of the S-matrix, taking into account the fact that in Lorentzian signature the
path integral is weighted by $e^{iS}$. 
On the other hand, in the notation of \cite{polchinski}, the same amplitude would have
been given by
\be\label{epolgrav}
i\, g_c^3 \, {8\pi\over g_c^2}\, \langle V_1(z_1) V_2(z_2) V_3(z_3)\rangle= 8\pi\, i\, g_c\,
\langle V_1(z_1) V_2(z_2) V_3(z_3)\rangle, \qquad g_c={\kappa\over 2\pi}\, .
\ee
The factor of $g_c^3$ arises from the convention that each closed string vertex operator is 
accompanied by a factor of $g_c$ and the $8\pi/g_c^2$ factor multiplies every sphere 
amplitude, determined in \cite{polchinski} by the requirement of factorization. The relation 
$g_c=\kappa/(2\pi)$ was needed to get the correct three graviton coupling as computed
from \refb{eeff1}, \refb{eeff2}. 

We now see that \refb{egrav} and \refb{epolgrav} agree.
Since it was shown in \cite{polchinski} that \refb{epolgrav} 
computed with these vertex operators \refb{egravminone} 
agrees with the one computed from the
Einstein-Hilbert action with gravitational coupling $\kappa$, we conclude that
$\kappa$ appearing in \refb{eclosedaction} 
is the
gravitational coupling constant appearing in \refb{eeff1}. 
Once the conventions have been matched,
it follows that all the higher order amplitudes computed from the action \refb{eclosedaction} 
also agree with those computed in \cite{polchinski}. We shall now briefly indicate how this works for the
four point function. According to \refb{eclosedaction} there will be a contribution to the four
point function obtained by joining a pair of three point vertices by a propagator. In the
Euclidean theory
three point vertices are each proportional to $4\kappa$ times appropriate three point functions
on the sphere, while it follows from \refb{e130} that  the propagator is given by
\be\label{e137}
-{1\over 2} \, \bar b_0 b_0(L_0+\bar L_0)^{-1}\, \delta_{L_0,\bar L_0} 
= {1\over 2}\, {1\over 2\pi} \, 
b_0\, \bar b_0\int_0^\infty ds \int_0^{2\pi} d\theta \, e^{-s(L_0+\bar L_0)}
e^{i\theta(L_0-\bar L_0)}\, .
\ee
Standard manipulation in conformal field theory now 
shows that the effect of the exponential factors and sum over all the
internal states in the propagator is to sew the two three punctured spheres into a four punctured 
sphere. The $b_0$, $\bar b_0$ factors convert one of the unintegrated vertex operators into
an integrated vertex operator and the integral over $s$ and $\theta$ generates integration over the
location $z$ of the integrated vertex operator with measure $d^2z/2$ where for $z=x+iy$,
$d^2z\equiv 2dxdy$. This has been reviewed in appendix \ref{sa}. 
Therefore after Wick rotation to Lorentzian signature, we get a net normalization 
factor:
\be\label{efirst}
i\, (4\kappa)^2 \, \left({1\over 4\pi}\right)\, {1\over 2} = 2\, i\, \kappa^2/\pi\, ,
\ee
besides the integral over the sphere four point function with three fixed and one
integrated vertex operators with measure $d^2z$. On the other hand,
according to the prescription of \cite{polchinski} the amplitude will get a factor of $g_c^4$
from the four vertex operators, a factor of $8\pi/g_c^2$ from the sphere and the standard factor
of $i$ for the S-matrix. This generates a
multiplicative factor:
\be
i\, g_c^4 \times 8\pi\, g_c^{-2} = i\, 8\, \pi \, (\kappa/(2\pi))^2 = 2\, i\, \kappa^2/\pi\, .
\ee
This is in agreement with \refb{efirst}. 
This agreement is not surprising, since the 
normalization of the amplitude was fixed in \cite{polchinski} by demanding that the
amplitudes factorize correctly, while in the amplitudes computed from string field theory, 
the factorization of the amplitude is guaranteed.

This can also be generalized to higher point function. Given an $n$-point function, adding another
vertex operator can be achieved by sewing of a three point function using a propagator. 
From \refb{e3.4}, the three
point function gives a factor of $4\kappa$, whereas the propagator generates an integral with
measure $d^2 z/ (8\pi)$. Therefore the net effect is multiplication by a factor of
$\kappa/(2\pi)=g_c$ and the integration over the location of the puncture with measure $d^2z$.
This agrees with the prescription of \cite{polchinski}.

The Ramond sector of closed string field theory is somewhat more involved, but we shall not need
this for our analysis. 

\subsection{Open string sector of string  field theory} \label{s2.3}

We now turn to the open string sector of the open-closed string field theory on a D$p$-brane. 
The NS sector
string field $|\psi_{NS}\rangle$ is taken to be a state with picture number $-1$ in the open
string Hilbert space, and the quadratic and 
cubic terms in the action
take the form:
\be\label{eopenns}
{1\over g_o^2} \left[ {1\over 2}\langle \psi_{NS}|Q_B|\psi_{NS}\rangle +{1\over 3!}
\{\psi_{NS}^3\} \right]\, ,
\ee
where in the definition of $\{V_1V_2V_3\}$ we include disk amplitudes with one PCO insertion, 
computed with the standard normalization given in \refb{eopennorm} 
and sum over
{\it both cyclic ordering
of the open string vertex operators $V_1,V_2,V_3$}. 
This explains the factor of $1/3!$ instead of the usual factor of $1/3$. $g_o$
is the open string coupling whose relation to the closed string coupling constant $\kappa$ will be
given later. If we define the field 
$|\phi_{NS}\rangle$ via,
\be
|\psi_{NS}\rangle = g_o |\phi_{NS}\rangle\, ,
\ee
then, up to this order, the action takes the form
\be \label{eopenphi}
{1\over 2}\langle \phi_{NS}|Q_B|\phi_{NS}\rangle +{g_o\over 3!}
\{\phi_{NS}^3\} \, .
\ee
Since in the Siegel gauge $Q_B$ is replaced by $c_0L_0$, and since $L_0$ acting on open string
states  has an additive term
$k^2$, the kinetic term has standard normalization. Therefore the 3-point coupling between three
physical open string states is given by $g_0$ times the disk 3-point function of the vertex 
operators with the standard normalization \refb{eopennorm}, 
without any additional factor.  Furthermore, following analysis similar to the one 
described for closed strings,
one can show that each additional external open string 
state gives an additional factor of $g_o$, and the
new vertex operator has to be converted to integrated picture and integrated along the real axis.
This agrees with 
the normalization used in \cite{polchinski}. Therefore the $g_o$ appearing in \refb{eopenns} agrees
with the one used in \cite{polchinski}.\footnote{One should keep in mind however that the relation
between $g_o$ and $\kappa$ or $g_s$ depends on the value of $p$, i.e. the particular D$p$-brane
we are considering.}

The infinitesimal gauge transformation parameter of the NS sector of the classical open string field
theory corresponds to an arbitrary NS sector state $|\theta\rangle$ of ghost number 0.
The gauge transformation law up to order $\phi_{NS}$ takes the form:
\be \label{egaugeopen}
\delta |\phi_{NS}\rangle = Q_B|\theta\rangle - g_o\, [\theta \phi_{NS}]\, ,
\ee
where $[AB]$ is defined so that for any state $|C\rangle$,\footnote{When the states
$A$, $B$, $C$ are not all grassmann odd, the contributions to $\{ABC\}$ from different
cyclic orderings come with opposite signs\cite{1907.10632}, {\it e.g.} in Witten's open string
field theory\cite{wittensft1,wittensft2}, $[\theta\phi_{NS}]=\theta * \phi_{NS}-\phi_{NS}*\theta$ .}
\be
\langle C|[AB]\rangle = \{CAB\}\, .
\ee

Finally we turn to the Ramond sector of the theory. Usually the construction of the kinetic term
requires either adding a free field\cite{1703.06410} or 
including a projection operator\cite{1508.00366}. However the construction simplifies 
if we
focus on the  effective 
action involving only the zero mass level states, after integrating out all the massive modes.
In this case we can 
take the classical string field to be a state
$|\psi_R\rangle$
of the open string of ghost number 1 and picture number $-1/2$ and the action up to the
cubic order can be taken to be of the form:
\be \label{eraction}
S= {1\over g_o^2}\left[
{1\over 2} \langle \psi_{R} | \YY_0\, Q_B|\psi_{R}\rangle + {1\over 3!} \{ \psi_R^2 \psi_{NS}\}
\right], 
\ee
where $\{ \psi_R^2 \psi_{NS}\}$ is given by the disk amplitude without any PCO insertion and,
\be
\YY_0 =\ointop {dz\over z} \, \YY(z), \qquad \XX_0 = \ointop {dz\over z} \, \XX(z)\, .
\ee
The $\ointop$ includes a factor of $1/(2\pi)$ so that $\ointop {dz/ z}=1$. 
For the full string field theory this is not an acceptable action since the Hilbert space
contains states in the kernel of $\YY_0$, but at mass level zero this problem is
absent.
Defining 
$|\phi_R\rangle = |\psi_R\rangle/g_o$, we can express the action as
\be\label{erphi}
{1\over 2} \langle \phi_{R} | \YY_0\, Q_B|\phi_{R}\rangle + {g_o\over 3!} \{ \phi_R^2 \phi_{NS}\}
\,  .
\ee
Note that we have used the same coupling constant $g_o$ 
for the NS and R-sector action. This can be
seen from the fact that a four point amplitude of two NS and two R sector states has contribution
from a pair of R-R-NS interaction
vertices connected by an R-sector propagator and also one R-R-NS and one
NS-NS-NS interaction vertex connected by an NS sector propagator. Therefore if we use different coupling
constants for the R-R-NS and NS-NS-NS interaction terms, the moduli space integrands of these
two contributions to R-R-NS-NS amplitude will not match.

\subsection{Interaction between open and closed strings} 

We shall now describe the normalization of some interaction terms that involve
closed strings (and possibly open strings) on Riemann surfaces with boundaries. 
Since our analysis in \S\ref{s6}, where these interaction terms will be used, will involve 
product of four copies of the disk amplitude with identical interaction vertices, the overall sign
and factors of $i$ in these interaction terms will not be important and will be 
ignored.\footnote{If we want to be more careful, we need to include additional factor of $i$
in \refb{eopcl} in order to have compatibility with sewing relations. This is related to the fact that
for a disk amplitude with closed and open strings, if we make an $SL(2,R)$ transformation to
go from a configuration with one fixed closed string puncture and one fixed open string
puncture to one with three fixed open string punctures, the resulting integration measure over
the closed string puncture is given by $i \, d^2 z$ instead of $d^2z$.
}

The elementary
interaction term is the one point function of the closed string on the disk. The corresponding
term in the action, denoted by $\{\psi_c\}_D$ is defined via the relation:
\be\label{e148}
\{\psi_c\}_D = {\TT\over 2} \, \langle  (c_0^- \psi_c)\rangle_D
\ee
where $\TT$ is the tension of the D$p$-brane under consideration, and $\langle~\rangle_D$ on the
right hand side is the closed string one point function on the disk computed with the normalization
\refb{eopennorm}. The closed string is inserted at the center of the disk $z=0$
using the local coordinate $e^\beta\, z$, where $z$ is the coordinate system in which the disk is
described by $|z|\le 1$ and $\beta$ is a parameter that characterizes the string field theory under
consideration\cite{1907.10632}. \refb{e148} can be taken as the definition of the D-brane tension.
It has been shown in appendix \ref{sb} that  this definition of the brane tension agrees with 
the usual definition based
on the low energy effective action.

Next we shall describe the interaction term involving disk amplitudes with multiple insertions of 
closed strings and open strings. In the action it will appear as:
\be\label{eopcint}
\sum_{m,n} {1\over m!n!} \{\psi_c^m \psi_o^n\}_D=\sum_{m,n} {1\over m!n!} 
\kappa^m g_o^n \{\phi_c^m \phi_o^n\}_D\, ,
\ee
where $\psi_o$ stands for the open string fields $\psi_{NS}$ or $\psi_R$,
$\phi_c=\psi_c/\kappa$ and $\phi_o=\psi_o/g_o$ are the canonically normalized fields, and,
\be\label{eopcl}
\{\psi_c^m \psi_o^n\}_D =  
\pi \, \TT \, \int \langle  \psi_c^m \, \psi_o^n \, \rangle_D\, .
\ee
Here $\langle  \psi_c^m \, \psi_o^n \, \rangle_D$ denotes correlation function on the disk / 
upper half plane with appropriate number of PCO insertions, 
computed with the normalization \refb{eopennorm}, with the vertex operators
inserted with choice of local coordinates appropriate to the string field theory under consideration
and the integral runs over part of the moduli space of the associated Riemann surface with
punctures, as prescribed by the particular version of the string field theory we consider. 
If we use the SL(2,R) invariance to fix the position of one closed string puncture and one
open string puncture, then for the rest of the punctures the integration measure is fixed as
follows. 
For a variable closed string puncture at position $z=x+iy$, the integration measure is taken to be
$d^2z/(2\pi)$ where $d^2z=2dxdy$, 
whereas for a variable open string puncture at position $x$, the integration
measure is taken to be $dx$. This is consistent with the normalization of the integration measure
over closed string punctures on the sphere and open string punctures on the 
disk found in \S\ref{s2.2} and \S\ref{s2.3}.
The extra
factor of $2\pi$ in \refb{eopcl} 
relative to \refb{e148} reflects the fact that the disk with one closed string insertion
at the origin has a conformal Killing vector that rotates the disk around the origin, and the volume of
this group is $2\pi$. Therefore in the computation of the one point function of closed strings on the
disk there is an implicit division by a factor of $2\pi$ that needs to be removed in
\refb{eopcl}. 

With this normalizations, we can check iteratively that
the interaction terms will satisfy the appropriate sewing relations needed for the gauge invariance
of the theory.
For example,
let us consider a disk 
amplitude with $m$ on-shell closed strings and $n$ on-shell open strings with canonically
normalized external states.
Part of this contribution comes from a Feynman diagram where a 
closed string three point vertex with two external states is connected to a disk amplitude with
$m-1$ closed strings and $n$ open strings by a closed string propagator.
In this case we get a factor of $1/(4\pi)$ 
from the propagator \refb{e137}, and another factor of $1/2$ while writing $ds d\theta$
in terms of $d^2z$ as discussed above \refb{efirst} and in appendix \ref{sa}.
Therefore
the amplitude involves a factor of $4\kappa$ from the closed string three point vertex as given in 
\refb{egrav}, a factor of $\pi\TT\kappa^{m-1}g_o^n$
from the disk amplitude with $(m-1)$ closed string and $n$ open strings, a factor of
$(2\pi)^{-(m-2)}$ associated with the integration measure of 
the $(m-2)$ integrated closed string puncture
on the disk
and a factor of $1/(8\pi)$ from the closed string propagator. 
This gives a net factor of 
$\TT\kappa^m g_o^n/2\times (2\pi)^{-(m-2)}$ accompanying this diagram. 
On the other hand, the same amplitude also gets a contribution from the interaction vertex
\refb{eopcint} with $m$ external closed strings and $n$ external open strings, covering a different
region of the moduli space. 
The associated normalization factor is 
$\pi\TT \kappa^m g_o^n$ times $(2\pi)^{-(m-1)}$ since there are $(m-1)$ integrated
closed string punctures on the disk with $m$ closed string punctures. 
Therefore the two normalization factors match, as required by gauge invariance.
A similar analysis
involving sewing via an open string propagator connecting a disk amplitude with $m$ closed
strings and $(n-1)$ open strings and the disk amplitude with three open strings
can be used to check consistency of the relative normalization given in \refb{eopcl} for $(m,n)$ 
and $(m,n-1)$.

\subsection{Relation between the different coupling constants} \label{s3.4}

We are now in a position to discuss the relation between $\kappa$, $g_o$ and $\TT$. 
In the following we shall ignore factors of $i$ and minus signs in the intermediate steps
since $\kappa$, $g_o$ and $\TT$
are all positive.
The relation
between $g_o$ and $\TT$ may be found as follows. Let us consider a disk 
amplitude with $m$ on-shell closed strings and $n$ on-shell open strings with canonically
normalized external states.
Part of this contribution comes from 
the interaction vertex \refb{eopcl} with $m$ closed strings and $n$ open strings, with 
associated normalization factor $\pi\TT \kappa^m g_o^n$ times $(2\pi)^{-(m-1)}$.
We shall write this as $\TT \kappa^m g_o^n/2\times (2\pi)^{-(m-2)}$.
On the other hand,
the same amplitude receives contribution from another class of 
Feynman diagrams in which 
a disk amplitude with $p$ closed string states and $q$ open string states is 
joined to another disk amplitude with $m-p$ closed string states and $n-q+2$ open string
states by an open string propagator. 
In this case this amplitude gets a 
factor of $\pi\, \TT \kappa^p g_o^q \times (2\pi)^{-(p-1)}$ and $\pi\, \TT \kappa^{m-p} g_o^{n-q+2}
\times (2\pi)^{-(m-p-1)}$ from the
two interaction vertices. The
Siegel gauge open string propagator
\be
b_0 (L_0)^{-1} = b_0\int_0^\infty e^{-s\, L_0}\, ,
\ee
does not generate any extra factor. 
This gives a net factor of $\pi^2\TT^2 \kappa^m g_o^{n+2}\times (2\pi)^{-(m-2)}$. 
Equating the two factors associated with the two Feynman diagrams we get
$\TT\kappa^m g_o^n/2=\pi^2\TT^2 \kappa^m g_o^{n+2}$. This gives
\be
\TT = {1\over 2\pi^2 g_o^2}\, .
\ee
This agrees with the result of \cite{9911116} obtained by 
different method and also with the result of \cite{polchinski}. 
For D-instantons we shall
label $\TT$ as
\be \label{ettgs}
\TT = {2\pi\over g_s}\, .
\ee
Therefore, we have
\be\label{egogs}
g_o^2 = g_s / (4\pi^3)\, .
\ee
$g_s$ is a useful parameter since $\tau = a + i \, g_s^{-1}$, where $a$ is the vacuum expectation value
of the Ramond-Ramond scalar, transforms as $\tau \to -1/\tau$ under S-duality
transformation.

The relation between $\kappa$ and $\TT$ can be found by considering the annulus zero point
function. On the one hand, this can be obtained by joining a pair of disk one point function of
closed strings by a closed string propagator. Since the  disk one point function of canonically
normalized closed string is proportional to $\kappa\TT$, this contribution will be proportional to
$(\kappa\TT)^2$. On the other hand this contribution may be expressed as an integral of the open
string partition function that does not depend on any parameter. Equating these two expressions
we can determine
$\kappa \TT$. This computation was carried out in \cite{polchinski} and since our conventions
for the parameters agree with that of \cite{polchinski} we just state the result:
\be\label{ekatt}
\kappa^2 \TT^2={1\over 2}\, (2\pi)^{7-2p} \, .
\ee
For D-instantons $p=-1$ and $\TT=2\pi/g_s$. This gives
\be\label{egska}
\kappa^2 = 2^6 \, \pi^7\, g_s^2\, .
\ee

\sectiono{Normalization of the D-instanton amplitude} \label{s4}

The general expression for the contribution to an amplitude due to a single D-instanton in
type IIB string theory, with action $2\pi/g_s$, takes the form
\be \label{eanrel}
\NN\, e^{-2\pi/g_s}\, \AAA\, ,
\ee
where $\NN$ is a normalization
constant and $\AAA$ is the usual world-sheet contribution to the amplitude. We have not explicitly
written down the $e^{2\pi i a}$ factor since we have not switched on RR scalar background, but the
presence of this factor follows from general considerations.
Our goal in this section will be to compute $\NN$. As mentioned below \refb{efinresult}, we shall 
not be careful about the overall phase of $\NN$ since it can be absorbed into a shift of $a$.

\subsection{Annulus partition function}

The general procedure for computing the normalization of the D-instanton amplitude was 
described in \cite{2101.08566}. As in \cite{2101.08566}, 
we shall formally write the normalization $\NN$ as:
\be \label{e4242}
\NN=i\, \zeta\, e^A\,  .
\ee
Here  $\zeta$ is a possible multiplier factor that specifies what multiple of the
full steepest
descent contour of the D-instanton is included in
the actual integration contour over the string fields. This will be
analyzed in \S\ref{s5}.
The factor of $i$ is common to all string amplitudes and
reflects the usual factor of $i$ that appears while relating the analytic continuation of the
Euclidean momentum space Green's functions to the S-matrix via the LSZ prescription. 
$A$ is the annulus partition function, formally written as\cite{polchinski}
\be \label{ebreak}
A=\int_0^\infty {dt\over 2t}  \, \left[{1\over 2}\,  \eta(it)^{-12}\left\{\vt_3(0|it)^4-\vt_4(0|it)^4
-\vt_2(0|it)^4 +\vt_1(0|it)^4
\right\}\right]\, ,
\ee
where the $\vt_i$'s are the Jacobi theta functions and $\eta$ is the Dedekind $\eta$ function.
The coefficient of $e^{-2\pi \, n\, t}$ inside the square bracket counts the difference between the
bosonic and fermionic open string states on the
D-instanton with $L_0$ eigenvalue $n$. The first two terms inside the square bracket reflect the
contribution from the NS sector states and the last two terms reflect the contribution from the
R sector states. The last term is actually zero, but we have written it here since this is the form
in which it arises when we take the trace over open string states. The $1/2$ inside the
square bracket comes from the GSO projection operator $(1+(-1)^f)/2$ where $f$ is the
world-sheet fermion number.
 
Now the annulus partition function 
$A$ given in \refb{ebreak}
actually vanishes due to cancellation between the NS and R sector states. 
However this cancellation
cannot be trusted since the $L_0=0$ sector represents 
NS and R sector zero modes for which
\refb{ebreak} is not applicable. Nevertheless the cancellation in the $L_0>0$ sector shows that
the contribution to $\NN$ comes entirely from the zero mode sector. Our strategy, following
\cite{2101.08566}, will be to 
represent the zero mode contribution to $\NN$ as integrals over the zero mode 
string fields, and then explicitly carry out these integrals.

To proceed further, it will be useful to regulate the contribution from the $L_0=0$ states 
to \refb{ebreak} by giving a small positive value to $L_0$. This can be achieved for example
by considering open strings stretched from one D-instanton to a neighboring D-instanton
separated by a small distance $a$ and noting that in the limit of zero separation the spectrum
reduces to that of open strings with two ends lying on the same D-instanton. For 
non-zero separation between the two D-instantons,
both the NS and the R-sector modes get a small positive contribution to $L_0$ given by
$h=a^2/(4\pi^2)$, introducing an additional multiplicative factor $e^{-2\pi t h}$ in the
integrand. Noting that the term inside the square bracket in \refb{ebreak} 
gets a contribution of 8 each
from the NS and the R-sector zero modes,  
we can express the regulated zero mode contribution to \refb{e4242} as:
\be
\NN = i\, \zeta \, \exp\left[ \int_0^\infty {dt\over 2t} \left(8\, e^{-2\pi t h} - 8\, e^{-2\pi t h}\right)\right]\, .
\ee
We can now use the general result\footnote{To arrive at \refb{egeneral} we need to put a lower
cut-off $\eps$ on the $t$ integral and take the $\eps\to 0$ limit at the end of the calculation.
A discussion on this may be found in \S\ref{s8} and \cite{2012.00041}.}
\be\label{egeneral}
\int_0^\infty {dt\over 2t} \, \left[\sum_{i=1}^n e^{-2\pi t h_i^b} - \sum_{i=1}^n e^{-2\pi t h_i^f}\right]
={1\over 2} \ln {\prod_{i=1}^n h_i^f\over \prod_{i=1}^n h_i^b}\, ,
\ee
to express $\NN$ as
\be 
\NN =  i\, \zeta \,  \sqrt{h^8\over h^8}\, .
\ee
For reasons that will be clear soon, we shall express this as an integral of the form:
\be\label{estrange}
\NN=i\, \zeta\, \int \left\{\prod_{\mu=0}^9 {d\xi_\mu\over \sqrt{2\pi}}\right\}\,  dp\,  dq\, 
\exp\left[-{1\over 2} h \sum_{\mu=0}^9 \xi_\mu \xi^\mu - h \, p\, q\right]
\int \prod_{\alpha=1}^{16}
d\chi_\alpha \, \exp\left[{1\over 2}g_{\alpha\beta} \chi_\alpha \chi_\beta
\right]\, ,
\ee
where $\xi_\mu$ are grassmann even modes, $p,q$ are grassmann odd modes,
$\chi_\alpha$ are grassmann odd modes and $g_{\alpha\beta}$ is an anti-symmetric,
$16\times 16$ hermitian matrix with the property:
\be
g^2 = h\, I_{16}\, ,
\ee
where $I_{16}$ is the $16\times 16$ identity matrix. Note that even though we have 
written the quadratic term in $\xi^\mu$ as $\xi_\mu\xi^\mu$, in euclidean signature this
is just $\sum_\mu(\xi^\mu)^2$ and the integral over the $\xi^\mu$'s is well-defined.

We shall now proceed as follows.
\begin{enumerate}
\item First we shall show that \refb{estrange} may be interpreted 
as the Siegel gauge fixed path integral of the
open string field theory on the D-instanton with appropriate normalization. Up to
normalization, the modes $\xi^\mu$ 
will represent the translation modes of the D-instanton in the $h\to 0$ limit, 
the modes $p$ and $q$ will
represent Faddeev-Popov ghosts in the NS sector and the modes $\chi_\alpha$ will 
represent the fermionic collective modes on the D-instanton in the $h\to 0$ limit.
\item Then we shall show that the Siegel gauge becomes singular in the $h\to 0$ limit,
and this is the reason why the coefficient of the $p\, q$ term, representing the ghost kinetic
operator, vanishes. The remedy will be to work with the original gauge invariant
path integral before gauge fixing.
\item We shall integrate over the collective modes at the end following standard procedure.
In particular we shall determine the 
correct normalization factor that relates the modes $\xi^\mu$ to the locations $\wt\xi^\mu$ 
of the D-instanton in Euclidean space time. The integration over the $\wt\xi^\mu$'s will
then generate the standard energy momentum conserving delta function for the momenta of
external states entering the amplitude $\AAA$ in \refb{eanrel}. 
The integration over the modes $\chi_\alpha$ will force us to insert
the vertex operators of each of the sixteen $\chi_\alpha$'s into the world-sheet 
defining the amplitude $\AAA$, since
otherwise the integral will vanish.
\end{enumerate}

\subsection{Gauge invariant string field theory in the $L_0=0$ sector}

Since open strings living on the D-instanton do not carry any continuous momenta, the
associated open string field theory is zero dimensional, containing a discrete set of modes.
Since we shall be working with only the $L_0=0$ sector, we begin by listing
the basis states in this sector.\footnote{Note that in the regulated version, what we refer
to as $L_0=0$ states actually have $L_0=h$.} 
\ben \label{ensstates}
\hbox{NS} &:& 
 \beta_{-1/2} c_1 |-1\rangle, \quad
c_1 d^\mu_{-1/2} |-1\rangle, \quad \beta_{-1/2} c_0 c_1  |-1\rangle,   \nonumber \\ && 
 \gamma_{-1/2} c_1 |-1\rangle, \quad
c_0 c_1 d^\mu_{-1/2} |-1\rangle,  \quad   \gamma_{-1/2} c_0 c_1 |-1\rangle,
\een
\be\label{errstates}
\hbox{R} \ :\ (\gamma_0)^n c_1 |-1/2,\alpha\rangle, \quad (\gamma_0)^n c_0 c_1 |-1/2,\alpha\rangle  \, ,
\ee
where we have defined,
\be
|-1\rangle \equiv e^{-\phi}(0)|0\rangle, \qquad  |-1/2,\alpha\rangle
= e^{-\phi/2} S_\alpha(0)|0\rangle\, .
\ee
Since classical open string fields carry ghost number 1, we have the following expansion of the
classical
fields $|\phi_{NS}\rangle$ and $|\phi_R\rangle$ introduced in \refb{s2.3}:
\be\label{eexpandsusy}
|\phi_{NS}\rangle =
i\, \phi^1 \, \beta_{-1/2} c_0 c_1  |-1\rangle 
+ \xi_\mu  c_1 d^\mu_{-1/2} |-1\rangle \, ,
\ee
\be \label{errexpand}
|\phi_R\rangle = \chi_\alpha c_1 |-1/2,\alpha\rangle\, ,
\ee
where $\phi^1$ and $\xi_\mu$ are grassmann even modes and $\chi_\alpha$ for 
$1\le \alpha\le 16$ are grassmann odd
modes. The $i$ multiplying the coefficient $\phi^1$ reflects the fact that the reality
condition on the string field requires an expansion of the form given in \refb{eexpandsusy}
with real $\phi^1$\cite{9705038}.
As discussed in \S\ref{s2.3}, the kinetic term of the action is given by:
\be\label{ekinetic}
S=S_{NS}+S_R, \quad S_{NS}={1\over 2} \langle \phi_{NS} | Q_B|\phi_{NS}\rangle,
\quad S_R=
{1\over 2} \langle \phi_{R} | \YY_0\, Q_B|\phi_{R}\rangle\, .
\ee
Since gauge transformation parameters are described by states of ghost number 0, we
see from \refb{errstates} that there is no gauge transformation parameter in the R sector, while the
NS sector contains a single gauge transformation parameters $\theta$: 
\be \label{egaugepar}
|\theta_{NS}\rangle = i\, \theta \,  \beta_{-1/2} c_1 |-1\rangle\, ,
\ee
where again the factor of $i$ reflects that real gauge transformation parameter in string field
theory corresponds to real $\theta$.
Classical gauge transformation law
\be
\delta |\phi_{NS}\rangle = Q_B|\theta_{NS}\rangle\, ,
\ee
translates to the following transformation of $\phi^1$ and $\xi^\mu$:
\be\label{etrspsi}
\delta\phi^1 =  \theta\, \langle -1|\gamma_{1/2} c_{-1} Q_B \beta_{-1/2} c_1 |-1\rangle\,,
\qquad \delta \xi^\mu = i\, \theta\, \langle -1| d^\mu_1 c_{-1} c_0 Q_B  \beta_{-1/2} c_1 |-1\rangle\, .
\ee
The partition function of the theory may now be
defined as:
\be\label{edefpart}
I = \int \left\{\prod_{\mu=0}^9 d\xi^\mu \right\} d\phi^1 \left\{\prod_{\alpha=1}^{16} 
d\chi_\alpha\right\} e^S
\Bigg/ \int d\theta\, .
\ee
At this stage the overall normalization of the partition function has been chosen arbitrarily. The
final result will be independent of this choice.

\subsection{Gauge fixing to Siegel gauge}

We now consider the
Siegel gauge in the NS sector: 
\be
b_0|\phi_{NS}\rangle=0\, .
\ee
This translates to:
\be
\phi^1=0\, .
\ee
Using \refb{etrspsi} we see that the corresponding Faddeev-Popov determinant is given by:
\be \label{efpdet}
\langle -1|\gamma_{1/2} c_{-1} Q_B \beta_{-1/2} c_1 |-1\rangle\, .
\ee
This may be represented by introducing
a pair of grassmann odd ghost fields $p,q$ defined via,
\be\label{eghf}
|\phi_{ghost}\rangle = -p\,  \gamma_{-1/2} c_1 |-1\rangle + 
q\, \beta_{-1/2} c_1 |-1\rangle\, ,
\ee
with action,
\be
S_{ghost} = {1\over 2} \langle \phi_{ghost}|Q_B|\phi_{ghost}\rangle = -p\, q\, 
\langle -1|\gamma_{1/2} c_{-1} Q_B \beta_{-1/2} c_1 |-1\rangle\, ,
\ee
so that $\int dp\, dq\,  e^{S_{ghost}}$ gives us the Faddeev-Popov determinant \refb{efpdet}. 

Since the Siegel gauge NS sector field and the ghost field  \refb{eghf}
both satisfy the Siegel gauge condition $b_0|\psi\rangle=0$, the BRST operator $Q_B$ reduces
to $c_0L_0$. Therefore the sum of the classical action and the ghost action now takes the form:
\be
S_{NS}+S_{ghost} 
= -{1\over 2} \sum_{\mu=0}^9 h\, \xi^\mu \xi_\mu -  h\, p\, q\, ,
\ee
where we have used the fact that the $L_0$ eigenvalues of these states is given by
$h$. On the other hand, using the form of $\YY$ given in \refb{eydef}, and of $Q_B$ given in
\refb{ebrs1}, \refb{ebrstcurrent},
and the fact that we need total $\phi$-charge $-2$ to get a non-vanishing disk
correlation function, we see that the Ramond action $S_R$ given in \refb{ekinetic} gets 
contribution only from the $\gamma T_F=\eta\, e^\phi\, T_F$ term in $Q_B$. Using 
the expansion \refb{evira} we may express the R sector kinetic term as
\be\label{egdef}
S_R={1\over 2} \, g_{\alpha\beta}\, \chi_\alpha \, \chi_\beta\, ,
\qquad g_{\alpha\beta}= \langle -1/2,\alpha| c_{-1} c_0 \, G_0^{(m)}\, c_1 |-3/2,\beta\rangle\, .
\ee

There is a comment that is in order here. If we regularize the path integral over zero modes by
considering open strings stretched between a pair of D-instantons, then the string field theory
action naturally pairs strings of opposite orientation. This necessarily doubles the spectrum
of the theory. In the NS sector we can avoid this problem by working with states with
Chan-Paton factors $\sigma_1$ or $\sigma_2$, since $Q_B$ will not mix these sectors.
However for the fermions, the operator $G_0^{(m)}$ will still pair the states in these two sectors, since
$G_0^{(m)}$ is linear in the perturbation that separates the D-instantons and this perturbation is
proportional to $\sigma_3$. Therefore if we just pick states in the sector $\sigma_1$ or
$\sigma_2$ then the kinetic term will vanish.
This can be avoided as follows. Let us suppose that we have 
separated the instantons along
the $x^1$ direction. In that case it follows from \refb{emattertdef} and
\refb{evira} that acting on the R sector ground state,
$G^{(m)}_0$ will be proportional to $\gamma^1$, which has non-zero matrix element between the
dotted and undotted spinors of the SO(8) group that acts on the coordinates $1,\cdots,8$.
If we now pick the dotted spinors of $SO(8)$ from the sector with Chan-Paton factor $\sigma_1$
and the undotted spinors of $SO(8)$ from the sector with Chan-Paton factor $\sigma_2$, then
$G^{(m)}_0$ will have non-zero matrix element between these states and will provide an
action of the form given in \refb{egdef}. This of course leaves the phase of the partition function
ambiguous, but as mentioned below \refb{efinresult}, this phase 
can be absorbed into a redefinition of the
vacuum expectation value $a$ of the RR scalar field.

Now using \refb{evira} and \refb{etope} we see that
\be\label{egalg}
\{ G^{(m)}_0, G^{(m)}_0\} = 2 \left(L^{(m)}_0 -{5\over 8}\right)\, .
\ee
On the other hand $c_1|-3/2,\beta\rangle$ has $L_0^{ghost}=-5/8$. Therefore \refb{egalg} gives,
\be
(G^{(m)}_0)^2 c_1 |-3/2,\beta\rangle = L_0 \, c_1 |-3/2,\beta\rangle= h\, c_1 |-3/2,\beta\rangle \, .
\ee
This in turn shows that the  matrix $g_{\alpha\beta}$ defined in \refb{egdef} squares to $h$ times
the $16\times 16$ identity matrix.

After gauge fixing, the partition function $I$ defined in \refb{edefpart} takes the form:
\ben\label{epartgf}
I &=& \int \left\{\prod_\mu d\xi^\mu \right\} dp\, dq\,  \left\{\prod_\alpha d\chi_\alpha\right\} e^{S+S_{ghost}} \nonumber \\
&=& \int \left\{\prod_\mu d\xi^\mu \right\} dp\, dq\,  \left\{\prod_\alpha d\chi_\alpha\right\} 
e^{-{1\over 2} \sum_{\mu=0}^9 h\, \xi^\mu \xi_\mu -  h\, p\, q +{1\over 2} g_{\alpha\beta}\, \chi_\alpha \, \chi_\beta}\, .
\een
Comparing this with \refb{estrange}, we get,
\be\label{ennirel}
\NN=i\, \zeta\, (2\pi)^{-5}\, I\, .
\ee

Let us now set $h=0$. Since $g_{\alpha\beta}$ squares to $h$ times the identity matrix,
$g_{\alpha\beta}$ also vanishes. In this case the action vanishes identically and the integrand becomes
independent of $\{\xi_\mu\}$, $\{\chi_\alpha\}$ and $p,q$. Lack of dependence on
$\{\xi_\mu\}$ and $\{\chi_\alpha\}$ may be traced to the fact that these are bosonic and
fermionic collective modes of the D-instanton, but the lack of dependence on $p$ and $q$ indicates
the vanishing of the Faddeev-Popov determinant and therefore the breakdown of the Siegel
gauge choice. 

\subsection{Gauge invariant partition function}

We circumvent the problem of breakdown of Siegel gauge
by replacing $I$ in \refb{ennirel} by the original
gauge invariant expression \refb{edefpart}. This gives,
\be\label{enewnn}
 \NN=i\, \zeta\, (2\pi)^{-5}\, \int \left\{\prod_\mu d\xi^\mu \right\} d\phi^1 \left\{\prod_\alpha d\chi_\alpha\right\} e^S
\Bigg/ \int d\theta\, .
\ee
We shall now set $h=0$ and regard $\xi^\mu$, $\phi^1$ and $\chi_\alpha$ as degrees of
freedom of the open string with both ends lying on the same D-instanton.
Substituting \refb{eexpandsusy} and \refb{errexpand} into \refb{ekinetic}, 
we get
\be 
S = -{1\over 4} (\phi^1)^2 \, .
\ee
We can now carry out the $\phi^1$ integral, generating a factor of $2\sqrt\pi$. This gives,
\be\label{enewernn}
 \NN=i\, \zeta\, (2\pi)^{-5}\, 2\sqrt \pi\, 
 \int \left\{\prod_\mu d\xi^\mu \right\}  \left\{\prod_\alpha d\chi_\alpha\right\} 
\Bigg/ \int d\theta\, .
\ee
It is to be understood that even though we have written the $\xi^\mu$ and
$\chi_\alpha$ integrals as part of $\NN$, these integrals need to be performed after
taking the product of $\NN$ with the rest of the world-sheet amplitude $\AAA$ 
appearing in \refb{eanrel}.

Our next task is to find the relation between $\xi^\mu$ and the D-instanton locations
$\wt\xi^\mu$ along the Euclidean space-time. This analysis proceeds as in \cite{2101.08566}. 
We note
that the integrated, zero picture vertex operators associated with the mode $\xi^\mu$
is given by
\be\label{evertzero}
\XX (z) \, i\, \sqrt 2\, \psi^\mu\, e^{-\phi} (z) = i\sqrt 2 \, \p X^\mu(z)\, . 
\ee
Now consider the effect of inserting the field $\xi^\mu$ into a disk amplitude of open and
closed strings with the closed strings carrying total momentum $p^\mu$. It follows from 
\refb{eopcl} and \refb{evertzero} that this will insert a vertex operator
\be
\int dz\, g_o\, \xi_\mu \,  i\sqrt 2 \, \p X^\mu(z)\, ,
\ee
with the integral running along the boundary of the disk. The factor of $g_o$ arises from the relation
$|\psi_o\rangle =g_o|\phi_o\rangle$ and that $\xi_\mu$ appears in \refb{eexpandsusy}
as coefficients in the expansion
of field $|\phi_{NS}\rangle$ with 
canonically normalized kinetic term.  Using the operator product expansion
\be
\p X^\mu(z) \, e^{ip_i.X(z_i)} = -{i\, p_i^\mu\over 2(z-z_i)}\, e^{ip_i.X(z_i)}\, ,
\ee
we can now evaluate the integration over $z$ and get a factor of
\be\label{ecomp1}
g_o\, \xi_\mu \, i\sqrt 2\, 2\, \pi \, i\, \left (-{i\over 2} \sum_i p_i^\mu\right)
= i\, g_o\, \pi\, \sqrt 2\, \xi_\mu \, \left( \sum_i p_i^\mu\right)\, ,
\ee
multiplying the original amplitude without $\xi^\mu$ insertion. 
On the other hand if $\wt\xi^\mu$ denotes the D-instanton location, then the dependence on
$\wt\xi^\mu$ of the amplitude is expected to be via a multiplicative factor of the form,
\be\label{ecomp2}
e^{i\, \wt\xi_\mu \left( \sum_i p_i^\mu\right)} = 1 + i \, \wt\xi_\mu \left( \sum_i p_i^\mu\right)+\cdots\, .
\ee
Comparing \refb{ecomp1} with \refb{ecomp2} we get,
\be
g_o\, \pi\, \sqrt 2\, \xi_\mu = \wt\xi_\mu\, .
\ee
This gives
\be
\prod_{\mu=0}^9 d\xi_\mu = g_o^{-10} \, \pi^{-10}\, 2^{-5}\, \prod_{\mu=0}^9 d\wt\xi_\mu \, .
\ee
By virtue of \refb{ecomp2}, the integration over $\{\wt\xi_\mu\}$ will generate the momentum
conserving delta function $(2\pi)^{10} \delta^{(10)}\left( \sum_p p_i\right)$, keeping in mind
that these integrals have to be performed after multiplying $\NN$ by the rest of the
world-sheet amplitude $\AAA$ as given in \refb{eanrel}.
Therefore, for now we leave the $\wt\xi^\mu$'s unintegrated and write
\be \label{enewnnone}
\NN =i\, \zeta\,  
g_o^{-10} \, \pi^{-10}\, 2^{-5}\, (2\pi)^{-5} \, 2\sqrt \pi\,  \int\left\{
\prod_\mu d\wt\xi_\mu\right\}\left\{ \prod_\alpha d\chi_\alpha\right\} 
\Bigg/ \int d\theta\, .
\ee

Next we shall analyze the result of integration over $\theta$. As in the case of \cite{2101.08566}, 
$\theta$ is 
related to the rigid gauge transformation parameter $\wt\theta$ under which an open string
connecting the D-instanton under study to a second spectator D-instanton picks up a factor
of $e^{i\wt\theta}$. Let us express the NS sector open string field $|\wh\phi_{NS}\rangle$
associated with the open string
connecting the two instantons by an expansion similar to \refb{eexpandsusy}, but with the
coefficients denoted by $\wh\xi^\mu$ and $\wh\phi^1$. This will carry a Chan-Paton factor 
$\pmatrix{0 & 1\cr 0 & 0}$. Then according to \refb{egaugeopen}, under the
gauge transformation generated by $\theta$, the transformation of $|\wh\phi_{NS}\rangle$ 
is given by,
\be
\delta|\wh\phi_{NS}\rangle=- g_o\, [\theta \wh\phi_{NS}]\, .
\ee
In particular the transformation law of $\wh\xi^\mu$ may be obtained by taking the
inner product of this with the state $c_1c_{0}  d^\mu_{-1}|-1\rangle $ with Chan-Paton factor 
$\pmatrix{0 & 0\cr 1 & 0}$. This gives, up to a sign,
\be
\delta \wh\xi^\mu = g_o\, \{ (c_1 c_{0}  d^\mu_{-1}|-1\rangle) 
( i\, \theta \,  \beta_{-1/2} c_1 |-1\rangle)
(\wh\xi_\nu c_1 d^\nu_{-1}|-1\rangle \}\, .
\ee
The trace over the Chan-Paton factors ensures that only one of the cyclic ordering 
contributes to the three point function on the disk that defines the $\{~\}$ in the above
equation. There is one PCO inside this correlation function. Taking
its location to coincide with the vertex operator $c\, \p\xi\, e^{-2\phi}$
of the state $\beta_{-1/2} c_1|-1\rangle$ multiplying the
gauge transformation parameter $\theta$, we
can convert the vertex operator of the gauge transformation parameter to:
\be
\XX(z)\, c \,  \p\xi\, e^{-2\phi}(z) = {1\over 2} I\, ,
\ee
where $I$ is the identity operator. Therefore we have
\be
\delta \wh\xi^\mu = g_o\, i\, \theta\, {1\over 2} \, \wh \xi_\nu \left\langle \left(i\sqrt 2 c\p c\,  \psi^\mu
e^{-\phi}(z_1)\right)
\left(i\sqrt 2 c \, \psi^\nu e^{-\phi}(z_2)\right)\right\rangle_D = {i\over 2} \, g_o\theta \, \wh \xi^\mu\, .
\ee
Comparing this with the infinitesimal rigid U(1) transformation $\delta \wh\xi^\mu =
i\wt\theta \xi^\mu$, we get $\theta = 2\wt\theta/g_o$. Since $\wt\theta$ has period $2\pi$, this
gives,
\be
\int d\theta = 4\pi/g_o\,.
\ee
Substituting this into \refb{enewnnone} we get
\be \label{enewnntwo}
\NN =i\, \zeta\,  
g_o^{-10} \, \pi^{-10}\, 2^{-5}\, (2\pi)^{-5} \, 2\sqrt \pi\,  {g_o\over 4\pi}\, \int\left\{
\prod_\mu d\wt\xi_\mu\right\}\left\{ \prod_\alpha d\chi_\alpha\right\} 
\, .
\ee
Finally note that the variables $\chi_\alpha$ are the coefficients of expansion of the field 
$|\phi_R\rangle$. It will be useful to express $\NN$ as integration over the coefficients of
expansion of the field $|\psi_R\rangle$ since this enters the interaction terms \refb{eopcl} without
any additional factor of $g_o$. To this end we introduce the variables $\wt\chi_\alpha$ via
\be \label{errexpandpsi}
|\psi_R\rangle = \wt\chi_\alpha\,  c_1 |-1/2,\alpha\rangle\, ,
\ee
Comparing this with \refb{errexpand} and using $|\psi_R\rangle = g_o |\phi_R\rangle$, we get
\be
\wt\chi_\alpha = g_o \, \chi_\alpha, \qquad \prod_\alpha d\chi_\alpha = g_o^{16}
\prod_\alpha d\wt\chi_\alpha\, ,
\ee
since $\chi_\alpha$ are grassmann odd variables. Substituting this into \refb{enewnntwo} we get
\be
\NN = i\, \zeta\, g_o^7\, 2^{-11} \, \pi^{-31/2} \, \int\left\{
\prod_\mu d\wt\xi_\mu\right\}\left\{ \prod_\alpha d\wt\chi_\alpha\right\} \, .
\ee
Finally we use \refb{egogs} to express this as:\footnote{This dependence on $g_s$ was first 
observed in \cite{bryappear}.}
\be\label{ennfin}
\NN= \NN_0 \, \, \int\left\{
\prod_\mu d\wt\xi_\mu\right\}\left\{ \prod_\alpha d\wt\chi_\alpha\right\} , 
\qquad 
\NN_0=i\, \zeta\, g_s^{7/2} \, 2^{-18} \, \pi^{-26}\, .
\ee
Integration over the grassmann odd variables shows that unless the rest of the amplitude
contains
insertions of the 16 $\wt\chi_\alpha$'s, the result vanishes identically. This will be discussed 
in \S\ref{s6}.

\sectiono{The multiplier factor} \label{s5}

There are two steps involved in 
the evaluation of the contribution due to a given instanton to the amplitude. The first is to
evaluate the contribution to the integral from the steepest descent contour / Lefschetz thimble
associated with each saddle point, including the classical vacuum and the various
instanton solutions. 
This amounts to integration over the full set of field fluctuations around each saddle point, with
each field integrated over its full range, but possibly deformed into the complex plane.
The second step is to express the 
actual integration contour, along which the path integral over the fields is to be performed,
as a (weighted) union of the Lefschetz thimbles for different 
saddle points\cite{1206.6272,1511.05977,1802.10441}. 
This associates a multiplier factor $\zeta$ to each instanton, with which we need to
multiply the steepest descent contribution, before we add the contribution to the amplitude.
This can sometime be non-trivial, {\it e.g.} in the analysis of \cite{2101.08566} 
in two dimensional bosonic
string theory, the multiplier factor associated with the D-instanton turned out to be 1/2.
Our analysis in \S\ref{s4} 
can be interpreted as part of the first step of the analysis since we
integrate all the modes from $-\infty$ to $\infty$ without worrying about whether the actual integration
contour involves the whole range. In this section we shall carry out the second step.

The D-instantons are complex solutions in the Euclidean type IIB string theory since the
RR scalar field is imaginary for the D-instanton solution\cite{9701093}. This may lead one to wonder whether
the D-instantons contribute to the amplitude at all, since usually the 
integration contour in the Euclidean
field theory runs over real field configurations, and therefore would
seem to miss the D-instanton configurations altogether. However we shall now argue that
this is not the correct way to view the D-instantons since they are not regular solutions of 
supergravity. Instead one should regard the D-instantons as regular solutions in the open string
field theory on an unstable D-brane system whose vacuum describes the regular perturbative 
vacuum\cite{0410103}.
For example a D-instanton in type IIB string theory can be regarded as a kink solution on a
non-BPS Euclidean D0-brane or a vortex solution in the euclidean D1-$\bar{\rm D}$1 brane
system. These are regular real solutions of the open string field theory and the reason that the
solution appears to be complex in the closed string theory is due to the fact that in the Euclidean
theory there is a complex contribution to the action of closed and open strings. For example in the
non-BPS D0-brane action there is a term proportional to $\int \chi\, dT$, where $\chi$ is the RR scalar
and $T$ is the open string tachyon, and in the Euclidean theory this gets a factor of $i$ due to
the $dT$ term acquiring an $i$ from the Wick rotated time direction. Since from the open string
perspective the D-instantons are real solutions, we conclude that the integration contour over the
open string fields include the full steepest descent contour of the D-instanton. Therefore the
multiplier factor $\zeta$ is 1.

\sectiono{4-graviton amplitude} \label{s6}

We shall now compute the leading D-instanton contribution to the four graviton amplitude.
Naively, the leading contribution comes from the product of four disk one point functions, with
a graviton vertex operator inserted at the center of each disk.
However the contribution from such
configurations to the four graviton amplitude vanishes due to the left over integration over the
$\wt\chi_\alpha$'s in \refb{ennfin}. The remedy is to consider a different amplitude where, besides
the four graviton vertex operators inserted at the centers of the four disks, we also have 
16 $\wt\chi_\alpha$'s as external states\cite{9701093}. 
As will be explained below, this gives a contribution to the effective action containing
product of 16 $\wt\chi_\alpha$'s and can give a non-zero result after integration over the
$\wt\chi_\alpha$'s. 

We shall now proceed as follows:
\begin{enumerate}
\item We shall first show that the disk amplitude with a single graviton and $n$ 
$\wt\chi_\alpha$'s vanish for $n=0,2$, so we need at least four $\wt\chi_\alpha$ insertions on the
disk to get a non-vanishing result. Therefore the 16 $\wt\chi_\alpha$'s must be equally distributed 
among the four disks.
\item Let $\AAA_{\alpha\beta\gamma\delta}(e, k) \, e^{ik.\wt\xi}$ be the 
disk amplitude of a single graviton of polarization
$e_{\mu\nu}$ and external open string modes $\wt\chi_\alpha$, $\wt\chi_\beta$, $\wt\chi_\gamma$
and $\wt\chi_\delta$. Note that we have included the dependence of the amplitude on the
position $\wt\xi$ of the instanton. 
This can be summarized by saying that the effective action of the open closed string field 
theory, after integrating out the $L_0>0$ modes, has a term\footnote{For writing the
effective action \refb{eeffact} we need an off-shell continuation of the function $\AAA_{\alpha\beta
\gamma\delta}$. Any off-shell continuation will serve our purpose since eventually we shall
evaluate this for on-shell external gravitons.}
\be \label{eeffact}
{1\over 4!} \, \int {d^{10}k\over (2\pi)^{10}} \, e^{ik.\wt\xi}\, \AAA_{\alpha\beta\gamma\delta}(h(k),k )\,\wt\chi_\alpha\wt\chi_\beta\wt\chi_\gamma
\wt\chi_\delta\, .
\ee
Eq.\refb{eanrel} and \refb{ennfin} now show that, 
after integrating out the open string modes, the closed string effective field theory
will have a term
\be
\NN_0\, e^{-2\pi/g_s}\, \int \prod_\mu d\wt\xi^\mu \ 
\prod_{\alpha=1}^{16} d\wt\chi_\alpha \, \exp\left[
{1\over 4!} \, \int {d^{10}k\over (2\pi)^{10}} \, e^{ik.\wt\xi}\, \AAA_{\alpha\beta\gamma\delta}(h(k),k)\, \wt\chi_\alpha\wt\chi_\beta\wt\chi_\gamma
\wt\chi_\delta\right]\, .
\ee
After expanding the exponential and using the result
\be
\int \prod_{\alpha=1}^{16} d\wt\chi_\alpha \, \wt\chi_{\alpha_1}\cdots \wt\chi_{\alpha_{16}}
=\eps_{\alpha_1\cdots \alpha_{16}}\, , \qquad
\int \prod_\mu d\wt\xi^\mu \, e^{i\wt\xi.\sum_i k_i} = (2\pi)^{10}\delta^{(10)}\left(\sum_i k_i\right)\, ,
\ee
we get a four graviton interaction term in the closed string effective field theory:
\ben 
&& 
\NN_0\, e^{-2\pi/g_s}\, {1\over 4!} {1\over (4!)^4} \, \eps_{\alpha_1\beta_1\gamma_1\delta_1
\cdots \alpha_{4}\beta_4\gamma_4\delta_4} 
\int {d^{10} k_1\over (2\pi)^{10}}\cdots {d^{10}k_4\over (2\pi)^{10}} \, (2\pi)^{10} \delta^{(10)}(k_1+k_2+k_3+k_4) \nonumber \\
&&\hskip 3in \times
\prod_{i=1}^4 \AAA_{\alpha_i\beta_i\gamma_i\delta_i}(h(k_i), k_i)\, .
\een
This generates the following contribution to the four graviton 
amplitude with polarizations
$e^{(i)}_{\mu\nu}$ and momentum $k_i$ with $1\le i\le 4$:
\be\label{e43}
\NN_0\, e^{-2\pi/g_s}\, {1\over (4!)^4} \, \eps_{\alpha_1\beta_1\gamma_1\delta_1
\cdots \alpha_{4}\beta_4\gamma_4\delta_4} 
\prod_{i=1}^4 \AAA_{\alpha_i\beta_i\gamma_i\delta_i}(e^{(i)}, k_i)\, 
(2\pi)^{10} \delta^{(10)}(k_1+k_2+k_3+k_4)\, .
\ee
Therefore our main task will be to compute $\AAA_{\alpha\beta\gamma\delta}(e,k)$.
\end{enumerate}

We shall begin by showing that the disk amplitude of
a single graviton with polarization
$e_{\mu\nu}$ vanishes. The vertex
operator in the $(-1,-1)$ picture up to a sign is 
$2 \, e_{\mu\nu} \, c\, \bar c \, e^{-\phi}e^{-\bar\phi}\psi^\mu 
\bar\psi^\nu$. We place the vertex operator at the point $i$ in the upper half plane and,
using the doubling trick, replace $\bar\psi^\nu(i)$ by $\psi^\nu(-i)$. The $\psi$ correlator now
produces a factor of $\eta^{\mu\nu}$ which  shows that the amplitude is proportional
to $\eta^{\mu\nu}e_{\mu\nu}$. This vanishes since the polarization tensor is traceless.

Next we compute the disk amplitude for one graviton and a pair of fermion zero modes
$\wt\chi_\alpha$ and $\wt\chi_\beta$. We insert the graviton vertex operator at $i$ on the
upper half plane as before, but convert this to $(0,-1)$ picture by taking the product with
the PCO $\XX$, 
represent $\chi_\alpha$ by the unintegrated $-1/2$ picture vertex operator 
$c\, e^{-\phi/2}\, S_\alpha$ inserted at the origin of the upper half plane, and
represent $\chi_\beta$ by an integrated $-1/2$ picture vertex operator $e^{-\phi/2}S_\beta(z)$
and integrate $z$ along the real axis. The amplitude is proportional to:
\be
\int dz\, \langle 2\, e_{\mu\nu} \, c\, \bar c\, \left\{\p X^{\mu} + i\, k_\rho\, \psi^\rho\psi^\mu\right\}
e^{ik.X}
e^{-\bar\phi}\bar\psi^\nu(i) \, c e^{-\phi/2}S_\alpha(0)\ e^{-\phi/2}S_\beta(z)\rangle_{UHP}\, .
\ee
We can now use the doubling trick to convert this to a correlation function on the 
full complex plane:
\be
\int dz\, \langle 2\, e_{\mu\nu} \, c \left\{\p X^{\mu} + i\, k_\rho\, \psi^\rho\psi^\mu\right\}
e^{ik.X}(i)\ 
c\, e^{-ik.X} e^{-\phi}\psi^\nu(-i) \, c e^{-\phi/2}S_\alpha(0)\ e^{-\phi/2}S_\beta(z)\rangle_{plane}\, ,
\ee
where all the fields are regarded as holomorphic.
The $z$ integral may be taken to pass either above or below the origin where $ce^{-\phi/2}S_\alpha$
is inserted since, according to \refb{espinope}, the difference 
between these two choices of contour is
proportional to $\gamma^\mu_{\alpha\beta}$ which is symmetric under the exchange of $\alpha$ and
$\beta$. Since eventually we need to contract this amplitude with $\eps_{\alpha\beta\cdots}$, the
contribution of this term will vanish. Let us take the contour to pass above the origin. We can now 
deform this to pick up the residue from the $\psi^\rho\psi^\mu$ insertion at $i$. Using
\refb{espinope} we see that the resulting
contribution will be proportional to
\be\label{evanshow}
{1\over 2}\, k_\rho \, e_{\mu\nu} \, (\gamma^{\rho\mu})_\beta^{~\gamma} \langle
c\,  e^{-\phi/2}S_\gamma e^{ik.X} (i) \
c\, e^{-ik.X} e^{-\phi} \psi^\nu(-i) \, c e^{-\phi/2}S_\alpha(0)\rangle_{plane}
\propto 
k_\rho \, e_{\mu\nu} \, (\gamma^{\rho\mu})_\beta^{~\gamma} (\gamma^\nu)_{\gamma\alpha}\,,
\ee
where $\gamma^{\mu_1\cdots \mu_n}$ is the totally antisymmetric product of 
$\gamma^{\mu_1},\cdots,\gamma^{\mu_n}$, normalized so that 
it is given by $\gamma^{\mu_1}\cdots \gamma^{\mu_n}$ when all the $\mu_i$'s are
different.
After expressing $\gamma^{\rho\mu}\gamma^\nu$ as a linear combination of 
$\gamma^{\mu\nu\rho}$, $\eta^{\rho\nu}\gamma^\mu$ and $\eta^{\mu\nu}\gamma^\rho$,
we see that \refb{evanshow}
vanishes using the symmetry and tracelessness of $e_{\mu\nu}$ and the 
condition $k^\mu e_{\mu\nu}=0$.

We shall now compute the amplitude $\AAA_{\alpha\beta\gamma\delta}(e,k)$ 
with one canonically normalized 
graviton and four $\wt\chi$'s inserted on the disk. During this computation 
we shall not be careful about factors of $i$ and minus signs since according to
\refb{e43} the result will be raised to fourth power. We shall convert the graviton
vertex operator at $i$ to an unintegrated 
zero picture vertex operator given in \refb{e125} and call this
$V_C$:
\be
V_C = 2\, e_{\mu\nu} \,  \bar c\, c\,\{\p X^\mu + i\, k_\rho\psi^\rho\psi^\mu\}
\{\bar\p X^\nu + i\, k_\sigma\bar\psi^\sigma\bar\psi^\nu\}\, e^{ik.X} +\cdots\, .
\ee
The $\cdots$ terms have non-zero $\phi$ charge and will not contribute to the correlation function.
We denote the unintegrated
$-1/2$ picture vertex operator of $\wt\chi_\alpha$ by $c\, W_\alpha$ where,
\be
W_\alpha = e^{-\phi/2} S_\alpha \, .
\ee
The corresponding integrated vertex operator is $W_\alpha$. We take the vertex operator of
$\wt\chi_\alpha$ to be unintegrated, placed at the origin, and those of 
$\wt\chi_\beta$, $\wt\chi_\gamma$ and $\wt\chi_\delta$
to be integrated along the real axis. Therefore, according to \refb{eopcl}, the
amplitude will be given by:\footnote{String field theory fixes the assignment of 
PCOs near
each degeneration. This translates to the following simple rule for the amplitude
under consideration. If the net number
of fermionic open string states that approach each other is even, then their picture number must add
up to $-1$, while if this is number is odd, then their picture number should add up to $-3/2$.
We can see that the picture number assignment we have taken is consistent with this rule when
two or three open strings come together, but when all four open strings come together, we need
to move one of the PCOs from the closed string vertex operator to near the
open string vertex operators. The effect of this movement can be computed using the trick of vertical integration following \cite{1408.0571}, 
and can be shown to vanish in this case. Therefore \refb{eorig} gives the correct
expression for the amplitude.}
\be \label{eorig}
\AAA_{\alpha\beta\gamma\delta}(e,k) =\kappa\, \pi \TT\,  \int dy_1 dy_2 d y_3 \, \langle V_C(i) \, c\, W_\alpha(0) W_\beta(y_1) W_\gamma(y_2)
W_\delta(y_3) \rangle_{UHP}\, .
\ee
The factor of $\kappa$ comes from having to express $\psi_c$ as $\kappa\phi_c$, since the
external graviton is taken to be the canonically normalized field.
Since $\wt\chi_\alpha$'s appear in the expansion of $|\psi_R\rangle$, it
follows from  \refb{eopcl} that we do not get any extra factor of $g_s$.

We can represent the vertex operator $V_C$ as
\be
V_C(z) = 2\,  e_{\mu\nu} \, \bar c(\bar z)\,  c(z)  \, U^\mu(z) \overline{U}^\nu(\bar z)\, ,
\ee
where $U_\mu$ is a holomorphic operator,
\be
U^\mu= (\p X^\mu + i\, k_\rho\psi^\rho\psi^\mu)\, e^{ik.X}\, .
\ee
This allows us to use the doubling trick and express the amplitude in terms of correlation functions
of holomorphic fields on the full complex plane:
\be
\AAA_{\alpha\beta\gamma\delta}(e,k)  = 2\, \pi\, \kappa\, \TT \,  e_{\mu\nu}\, 
\int dy_1 dy_2 d y_3 \, \langle  c\, U^\mu(i) \, c\, U^\nu(-i) 
\, c\, W_\alpha(0) W_\beta(y_1) W_\gamma(y_2)
W_\delta(y_3) \rangle\, ,
\ee
where it will be understood that due to Dirichlet boundary condition on $X^\mu$, 
the
$e^{ik.X}$ factor is replaced by $e^{-ik.X}$ in the expression for $U^\mu(-i)$.
Due to the symmetry arguments described earlier, the relative positions of the integration
contours does not matter.  We shall choose the
$y_1$ contour to be above the real axis and $y_2$ and $y_3$ contours to be below the real
axis with $Im(y_2)> Im(y_3)$. The holomorphic correlation functions will be normalized
following the open string prescription \refb{eopennorm}, with the $(2\pi)^{p+1}\delta^{(p+1)}(k)$
factor absent for D-instantons.

We can now deform the $y_1$ contour into the upper half plane and the $y_3$ contour into the
lower half plane to pick residues at $i$ and $-i$ respectively. For this we use the operator 
product expansion derived from \refb{espinope}:
\be
W_\alpha (y) \, U^\mu(z) = {1\over y-z} {i\over 4} \, k_\rho \,
(\gamma^{\rho\mu})_\alpha^{~\beta} 
W_\beta(z)\, e^{ik.X}(z)\, .
\ee
Similarly we close the $y_3$ contour in the lower half plane, picking up the residue
at $-i$.
This gives, after including the $(2\pi)^2$ factor from the residue theorem,
\ben
\AAA_{\alpha\beta\gamma\delta}(e,k)  &=& 8\, \pi^3\, \kappa\, \TT \,  e_{\mu\nu}\, 
\int dy_2 \, \Bigg\langle  c(i)\, {i\over 4} k_\rho (\gamma^{\rho\mu})_\beta^{~\beta'} \,
W_{\beta'}(i) \, e^{ik.X}(i)  \nonumber \\ && \hskip .5in 
c(-i)  {i\over 4} k_\sigma (\gamma^{\sigma\nu})_\delta^{~\delta'} 
W_{\delta'}(-i) \, e^{-ik.X}(-i) c(0) W_\alpha(0) \, W_\gamma(y_2)
\Bigg\rangle\, .
\een
Next we can deform the $y_2$ contour to pick the residue at $-i$ using the
operator product expansion  derived from \refb{espinope}:
\be
W_\gamma(y) W_{\delta'}(z)  = {1\over y-z} \, i\, (\gamma^\tau)_{\gamma\delta'} \, e^{-\phi}
\, \psi_\tau(z)\, .
\ee
This gives
\ben
\AAA_{\alpha\beta\gamma\delta}(e,k) &=& 16\, \pi^4\, \kappa\, \TT \,   e_{\mu\nu}\, 
\left\langle  c(i)\, {i\over 4} k_\rho (\gamma^{\rho\mu})_\beta^{~\beta'} 
W_{\beta'}(i)\, e^{ik.X}(i)\right. \nonumber \\
&&\hskip .5in  \left.   c(-i) {i\over 4} k_\sigma (\gamma^{\sigma\nu}\gamma^\tau)_{\delta\gamma} 
e^{-\phi}\psi_\tau(-i) e^{-ik.X}(-i) \, c(0) W_\alpha(0) 
\right\rangle\, .
\een
Finally we can use the result
\be
\langle c(i) W_{\beta'}(i) \, \, e^{ik.X}(i) c(0) W_\alpha(0) c(-i) e^{-\phi} \psi_\tau(-i)
\, e^{-ik.X}(-i)\rangle
= {i\over 2} (\gamma_\tau)_{\beta'\alpha}\, ,
\ee
and drop all factors of $i$ since we have not kept track of these factors even in the interaction
vertex \refb{eopcl} that we have been using. This gives,
\be
\AAA_{\alpha\beta\gamma\delta}(e,k) ={1\over 2}\,  \pi^4\, \kappa\, \TT \,  
e_{\mu\nu}\,  (\gamma^{\rho\mu}\gamma_\tau)_{\beta\alpha}
(\gamma^{\sigma\nu}\gamma^\tau)_{\delta\gamma} \, k_\rho\, k_\sigma\, .
\ee
Using the result,
\be
\gamma^{\rho\mu}\gamma^\tau=\gamma^{\rho\mu\tau} + \eta^{\mu\tau}\gamma^\rho
-\eta^{\rho\tau}\gamma^\mu\, ,
\ee
and the fact that $\gamma^\rho$, $\gamma^\mu$ are symmetric matrices and that
we eventually 
anti-symmetrize the amplitude
under the permutation of $\alpha,\beta$, $\gamma$, $\delta$, we can write
\be \label{eaafin}
\AAA_{\alpha\beta\gamma\delta}(e,k) = {1\over 2} \pi^4\, \kappa\, \TT 
\,   e_{\mu\nu}\,  (\gamma^{\rho\mu}_{~~~\tau})_{\beta\alpha}
(\gamma^{\sigma\nu\tau})_{\delta\gamma}  \, k_\rho\, k_\sigma\,.
\ee

Using \refb{e43} and \refb{eaafin}
we now get the single D-instanton contribution to the 
4-graviton amplitude:
\be
\NN_0\, e^{-2\pi/g_s}\, 
\left({1\over 2}\pi^4\, \kappa\, \TT\right)^4 \, {1\over (4!)^4}\, \eps^{\alpha_1\beta_1\gamma_1\delta_1\cdots
\alpha_4\beta_4\gamma_4\delta_4}\prod_{i=1}^4  \left\{
 e^{(i)}_{\mu_i\nu_i}\,  (\gamma^{\rho_i\mu_i}_{~~~~\tau_i})_{\beta_i\alpha_i}
(\gamma^{\sigma_i\nu_i\tau_i})_{\delta_i\gamma_i} \,  k^{(i)}_{\rho_i}\, k^{(i)}_{\sigma_i}
\right\}\, .
\ee
We now use the result:\footnote{This result was stated in \cite{9701093} up to an overall normalization 
factor. The proportionality between the two sides of \refb{e624} 
follows from space-time supersymmetry which fixes the
tensor structure of the four graviton amplitude.
We have computed the normalization by numerically evaluating both sides for special
cases.}
\be\label{e624}
\eps^{\alpha_1\beta_1\gamma_1\delta_1\cdots
\alpha_4\beta_4\gamma_4\delta_4}\prod_{i=1}^4  \left\{
 e^{(i)}_{\mu_i\nu_i}\,  (\gamma^{\rho_i\mu_i}_{~~~~\tau_i})_{\beta_i\alpha_i}
(\gamma^{\sigma_i\nu_i\tau_i})_{\delta_i\gamma_i} \,  k^{(i)}_{\rho_i}\, k^{(i)}_{\sigma_i}
\right\} = (4!)^4 \, 2^{12} \, K_c\, ,
\ee
where
\be\label{e625}
K_c(e_1,e_2,e_3,e_4)=t^{\mu_1\nu_1\cdots \mu_4\nu_4} \, t^{\rho_1\sigma_1\cdots \rho_4\sigma_4}\,
\prod_{j=1}^4 e^{(j)}_{\mu_j \rho_j}  k^{(j)}_{\nu_j} k^{(j)}_{\sigma_j}\, ,
\ee
and $ t^{\rho_1\sigma_1\cdots \rho_4\sigma_4}$ is defined via the relation:
\ben \label{e626}
&& t^{\mu_1\nu_1\cdots \mu_4\nu_4} \prod_{j=1}^4 f^{(j)}_{\mu_j} k^{(j)}_{\nu_j} 
= {1\over 8} \left[ 4 \, Tr(M_1M_2M_3M_4) - Tr(M_1M_2) Tr(M_3M_4)\right] + 
\hbox{2 permutations}\, ,
\nonumber \\
&& M_{i\mu\nu} \equiv k^{(i)}_{\mu} f^{(i)}_{\nu}- f^{(i)}_{\mu} k^{(i)}_{\nu}\, .
\een
This gives the amplitude to be
\be 
\NN_0\, e^{-2\pi/g_s}\, 2^8 \, (\pi^4\kappa\TT)^4\,  K_c\, .
\ee
Using \refb{ettgs}, \refb{egska} and \refb{ennfin}, and the result $\zeta=1$, 
we can express this as:
\be \label{efinr}
e^{-2\pi/g_s}\, i\,  g_s^{7/2} \, 2^{-18} \, \pi^{-26}\,  2^8
(\pi^4 \times  2^3 \pi^{7/2} g_s \times 2\pi/g_s)^4 K_c = i\,  e^{-2\pi/g_s} \, 2^{6}\,
\pi^{8}\, g_s^{7/2}\, K_c 
\, .
\ee
This reproduces \refb{efinresult}. We shall check in 
\S\ref{s7} that it agrees with the prediction of
S-duality.

\sectiono{Prediction for the D-instanton contribution to the four graviton 
amplitude from duality} \label{s7}

We shall now derive the prediction for the same amplitude using S-duality of type IIB
string theory. In the convention of \cite{polchinski}, which agrees with ours, the 
tree level scattering amplitude takes the form:
\be\label{etest}
{i\over 4} \, \kappa^2\, K_c\,  
\left[{64\over stu} + 2\zeta(3)\right]\, (2\pi)^{10} \, \delta^{(10)}(k_1+k_2+k_3+k_4)\, .
\ee
The first term can be identified as the contribution to the scattering amplitude from the
Einstein-Hilbert action\cite{sannan} and can be used to check the overall normalization of 
\refb{etest}. 
This is S-duality invariant by itself as can be seen by converting this result to the Einstein frame
by multiplying this be a factor of $1/g_s^2$. 
The second term can be identified
as the contribution from a new term in the action proportional to the fourth power of the 
Riemann tensor\cite{grosswitten}. 
This term is not S-duality invariant by itself, but admits a completion to an
S-duality invariant action by adding a one loop and non-perturbative terms\cite{9701093}. 
This modifies the
four graviton amplitude to:\footnote{The original paper\cite{9701093} had a typographical error in the
coefficient of the $e^{2\pi i\tau}$ term, but the correct coefficient can be found in later papers
{\it e.g.} in \cite{9704145}.}
\be
{i\over 4} \, \kappa^2\, K_c\,  
\left[{64\over stu} + 2\zeta(3) + {2\pi^2\over 3} \, g_s^2+ 4\, \pi \, g_s^{3/2}\,
\{e^{2\pi i\tau}+e^{-2\pi i\bar\tau}\}
+\cdots
\right]\, (2\pi)^{10} \, \delta^{(10)}(k_1+k_2+k_3+k_4)\, ,
\ee
where,
\be
\tau = {a} + {i\over g_s}\, ,
\ee
with $a$ being the expectation value of the RR scalar field. The coefficient of $e^{2\pi i\tau}$
gives the single D-instanton contribution to the amplitude and the coefficient of $e^{-2\pi i\bar
\tau}$
gives the anti-D-instanton contribution. Therefore
the expected contribution to the amplitude from a single D-instanton is:
\be
e^{2\pi i a}\, e^{-2\pi/g_s} \, {i\over 4} \, \kappa^2\, K_c\, 4\, \pi \, g_s^{3/2} = i\, e^{2\pi i a}\, 
e^{-2\pi/g_s}\, 
2^6 \, \pi^8 \, g_s^{7/2}\, K_c,
\ee
where in the last step we have used \refb{egska}. This agrees with \refb{efinr} for vanishing RR scalar.

\sectiono{Generalizations} \label{s8}

In this section we shall discuss possible generalizations of our analysis.

The computation of the D-instanton amplitude in this paper consisted of two parts. The subtle part
involved the computation of the normalization constant $\NN$ in \S\ref{s4}. This part of the computation
will be the same for all single D-instanton amplitudes in type IIB string theory, irrespective of the
number of external lines, their nature and the order of $g_s$ to which we want to compute the
amplitude. The second part of the analysis, that in \S\ref{s6}, is specific to the amplitude we are
interested in, and will have to be redone for a different amplitude.

Next we shall discuss generalization of this analysis to other theories.
Our analysis in this paper, as well as in \cite{2101.08566}, simplified since the contribution from the
$L_0>0$ states in the integrand of the annulus partition function \refb{ebreak}
cancelled and we had to deal with a finite number of modes. However this is
not necessary. 
Let us suppose that the
annulus partition function $A$ has the form:
\be\label{e8.1}
A=\int_0^\infty {dt\over 2t}\,  f(t)\, .
\ee
Then the key property that is needed to generalize 
our analysis is the vanishing of  $f(t)$
in the $t\to 0$ limit so that the integral does not have any divergence from the lower end. 
This is valid in any string theory without closed string tachyons, 
since the contribution from the small $t$ region 
can be interpreted as coming from the infrared region of a
single loop of closed strings emitted and absorbed by the 
D-instanton.  This contribution is finite as long as there are no closed string tachyons
and we have more than two non-compact
dimensions. For this reason  
we only have to deal with possible divergences from the $t\to\infty$ region associated
with the tachyonic and zero modes of the open string.
Let us express the tachyonic and the zero
mode contributions to $f(t)$ as
\be \label{e8.2}
\sum_{i=1}^m e^{-2\pi t h^b_i} - \sum_{j=1}^n e^{-2\pi t h_j^f}\, ,
\ee
where the sum runs over the non-positive $h_i^b$ and $h_j^f$ values. If $m$ and $n$ are
equal, then \refb{e8.2} vanishes at $t\to 0$, and 
we can analyze the contribution of \refb{e8.2} to $e^A$ by representing
it as integral over bosonic and fermionic modes as in this paper and deal with the zero modes
appropriately. For the tachyonic modes we can simply use the steepest descent contour as
in \cite{2101.08566}. For the rest of the contribution to $f(t)$, coming from $L_0>0$ modes,
we can evaluate the integral over $t$
in \refb{e8.1}
explicitly (if necessary numerically) and get a finite result since the integrand vanishes sufficiently
fast both as $t\to 0$ and $t\to\infty$. If on the other hand $m$ and $n$ in \refb{e8.2} are not
equal, we can simply include the contribution from appropriate number of positive $h_i^b$ or
$h_j^f$ 
modes in the sum in \refb{e8.2} to make them equal, and then proceed as before. One can
easily verify that the final result is independent of which set of positive
$h_i$ values  we include in
the sum in \refb{e8.2}.

This gives a systematic procedure for computing the contribution of a D-instanton to an 
amplitude from the steepest
descent contour (Lefschetz thimble) of the instanton. However we also need to understand
how the steepest descent contour fits inside the full integration contour. If the instanton is a
real solution in open string field theory on unstable D-brane system 
and has no tachyons, then we expect the full steepest
descent contour to be part of the integration contour and the multiplier factor will be unity.
Otherwise we need to do further analysis to evaluate the multiplier factor.

This shows that the ability to carry out systematic computation of D-instanton correction to
string theory amplitudes does not rely on supersymmetry but on the ultraviolet finiteness
of string theory. Finally we would like to note that the arguments given above hold also for
other Euclidean D-branes as long as they are wrapped on compact cycles and have more
than two transverse non-compact directions. Therefore the same method could be used to
compute the contribution to the superpotential induced by Euclidean D-branes in N=1
supersymmetric string compactification.

\bigskip

\noindent {\bf Acknowledgement:} I wish to thank Bruno Balthazar, Anirban Basu,
Rajesh Gopakumar, Michael Green, Sitender
Pratap Kashyap,
Victor Rodriguez, Jorge Russo, 
Congkao Wen, Xi Yin and Barton Zwiebach for useful discussions and Bogdan Stefanski for
critical comments on an earlier version of this manuscript. This work was
supported in part by the  Infosys chair professorship and the
J. C. Bose fellowship of 
the Department of Science and Technology, India. 

\appendix

\sectiono{Sphere four point function from sewing of two three point functions} \label{sa}

In this appendix we shall review how connecting a pair of three point interaction vertices by the
propagator \refb{e137} generates the four point amplitude with the normalization factor given
in \refb{efirst}. For simplicity we shall illustrate this in the context of bosonic string theory, but the
same analysis can be carried out for superstring theory.

Let us suppose that the three point interaction vertices are described by a three point function on the
sphere with vertex operators placed at 0, 1 and $\infty$. We shall denote by $z$ the global
coordinate on the complex plane and choose the local coordinate at 0 to be $z$ and that at $\infty$
to be $-1/z$. This choice is not symmetric under the permutation of the vertex operators, 
but will serve to
demonstrate the main point of the analysis, i.e. to determine the normalization given in
\refb{efirst}. We now denote the global coordinates associated with the two three
point vertices by $z$ and $z'$ and sew the puncture at 0 of the first interaction vertex with the
puncture at $\infty$ of the second interaction vertex using the sewing parameter
\be
q=e^{-s+i\theta}\, ,
\ee
where $s$ and $\theta$ are the parameters introduced in \S\ref{s2.2}. This gives
\be
z \, \left(-{1\over z'}\right) = q\, ,
\ee
i.e. $z=-qz'$. Therefore in the $z$ plane the punctures at $z'=0$ and $z'=1$ are located at 
$0$ and $-q$ respectively. The amplitude obtained by sewing two three point functions with the
propagator \refb{e137} is now given by:
\be
i (4\kappa)^2 \left({1\over 4\pi}\right)\,
 \int_0^\infty ds \int_0^{2\pi} d\theta \, \left\langle \bar c c W_1(1) \, \bar c
c W_2(\infty) 
\ointop dz z b(z) \ointop d\bar z \bar z \bar b(\bar z) 
\, \bar c cW_3(0) \, \bar c cW_4(-q)\right\rangle,
\ee
where $\ointop$ is a contour enclosing the points $0$ and $-q$ and $W_i$'s are dimension
(1,1) primaries in the matter sector. The $4\kappa$ factors come
from the three point functions as in \S\ref{s2.2} and the $i$ is the usual factor in the expression
for the S-matrix. We can now carry out the contour integrals to express this as:
\be
{4\, i \kappa^2 \over \pi} \int_0^\infty ds \int_0^{2\pi} d\theta \, \left\langle \bar c c
W_1(1) \, \bar c c W_2(\infty) \, 
\bar c  c W_3(0)\,  |q|^2 W_4(-q)\right\rangle\, .
\ee
Defining $w=-q=-e^{-s+i\theta}$ 
as the location of the fourth vertex operator and $d^2w=2dxdy$ for $w=x+iy$,
we can express this as
\be
{2\, i \, \kappa^2 \over \pi} \int d^2 w \, \left\langle \bar c c W_1(1) \, \bar c c W_2(\infty) 
\, \bar c cW_3(0)\, W_4(w)\right\rangle\, .
\ee
This reproduces \refb{efirst}.

\sectiono{Comparison of the two definitions of the brane tension} \label{sb}

In this appendix we shall check that the brane tension $\TT$ that enters \refb{e148}
agrees with the usual definition based
on the low energy effective action. For this we recall that the presence of a D$p$-brane 
gives a contribution to the action of the form:
\be\label{ebrane1}
-\TT\, \int d^{p+1}x \, \sqrt{-\det G}\, e^{-\Phi}\, ,
\ee
where $G$ denotes the string 
metric along the brane and $\Phi$ is the dilaton field. 
Using \refb{eeff2}, and assuming that the D$p$-brane
is placed at the origin of the transverse coordinates, \refb{ebrane1} leads to the following contribution
to the action linear in $h_{\mu\nu}$ and $\Phi$:
\be \label{ebrane2}
-\TT\, \int {d^{9-p}k_\perp\over (2\pi)^{9-p}} \, \left[
\kappa \sum_{\mu,\nu=0}^p \eta^{\mu\nu} h_{\mu\nu}
(0, k_\perp)  - \Phi(0,k_\perp) \right]\, ,
\ee
where $k_\perp$ denotes components of momenta transverse to the brane and we have
used the same symbol $\Phi$ to label the Fourier transform of the dilaton field. 

We can now compare 
this with the terms linear in $h_{\mu\nu}$ obtained from \refb{e148} using the expansion of 
$|\psi_c\rangle=\kappa|\phi_c\rangle$ from \refb{efieldexp}. 
Let us denote these coefficients by $h'_{\mu\nu}$ instead of the same symbol
$h_{\mu\nu}$ that appears in the expansion of the metric.
This gives a term in the action of the from:
\be
{1\over 2} \, \kappa\, \TT \, \int{d^{10}k\over (2\pi)^{10}}\, h'_{\mu\nu}(k)
\left\langle (-2\, c_0^- c\, \bar c\, e^{-\phi} \psi^\mu e^{-\bar\phi}  \bar \psi^\nu e^{ik.X}(0))\right\rangle_D\, ,
\ee
where 
\be
c_0^- = {1\over 2}\left( \ointop dw\, w^{-2} \, c(w) - \ointop d\bar w \, \bar w^{-2} \, \bar c(\bar w)\right)\, ,
\ee
with the contours evaluated around the origin of the disk and containing factors of 
$(\pm2\pi i)^{-1}$. One can map this into the correlation function
on the upper half plane by making appropriate 
transformation of coordinates and then use the doubling trick and
\refb{eopennorm}
to evaluate the matrix element. The result is,
\be \label{ebrane3}
{1\over 2} \, 
\kappa\, \TT\, \int {d^{9-p}k_\perp\over (2\pi)^{9-p}} \, \left[-\sum_{\mu=0}^p \eta^{\mu\nu} h'_{\mu\nu}
(0, k_\perp)  + \sum_{\mu=p+1}^9 \eta^{\mu\nu} h'_{\mu\nu}
(0, k_\perp)
\right]\, ,
\ee
where the relative minus sign between the two terms in the square bracket is due to the
difference in the boundary condition on the $\psi^\mu$'s for $\mu$ tangential and transverse 
to the D-brane. Similarly \refb{e148} can be used to calculate the term linear in the
scalar $\Psi$ that multiplies the state 
$(\beta_{-1/2}\bar\gamma_{-1/2}+\bar\beta_{-1/2}\gamma_{-1/2})c_1\bar c_1|-1,-1\rangle$ 
in the expansion of the string field. By choosing the 
normalization of $\Psi$ appropriately we can express this as:
\be \label{ebrane3a}
{1\over 2} \, 
\kappa\, \TT\, \int {d^{9-p}k_\perp\over (2\pi)^{9-p}} \, \Psi(k)\, .
\ee

In order to compare \refb{ebrane2} with the sum of \refb{ebrane3} and \refb{ebrane3a}, we 
need to know
the relation between the fields $(h_{\mu\nu},\Phi)$ and $(h'_{\mu\nu},\Psi)$.
Since $h_{\mu\nu}$ and $h'_{\mu\nu}$ are known to 
transform in the same way
under the linearized gauge transformation laws in supergravity and
closed string field theory respectively, they can differ at most by a term proportional to the
scalar field $\Phi$. Therefore
the general form of the relationship between the two sets of fields
takes the form:
\be \label{ehhprel}
h'_{\mu\nu} =h_{\mu\nu} + a \, \Phi\, \eta_{\mu\nu}, \qquad 
\Psi= b\, \Phi + c \, \sum_{\mu=0}^9 \eta^{\mu\nu} h_{\mu\nu}\, ,
\ee
for some constants $a$, $b$ and $c$. 
Note that for non-zero $c$, $\Psi$ transforms
under gauge transformation -- indeed this can be seen directly using the linearized gauge
transformation laws of closed string field theory.
We can find the constants $a$, $b$ and $c$ by  
comparing the action and gauge transformation laws 
of the low energy
supergravity with the action and gauge transformation laws of closed string field 
theory\cite{0506077},
but we
shall take a shortcut. 
Substituting \refb{ehhprel} into the sum of \refb{ebrane3} and \refb{ebrane3a}, we get:
\ben \label{ebrane22}
&& -{\kappa\TT\over 2}\, \int {d^{9-p}k_\perp\over (2\pi)^{9-p}} \, \Bigg[
 (1-c) \sum_{\mu=0}^p \eta^{\mu\nu} h_{\mu\nu}
(0, k_\perp) 
- (1+c) \sum_{\mu=p+1}^9 \eta^{\mu\nu} h_{\mu\nu}(0, k_\perp) \nonumber \\ &&
\hskip 2in + \left\{a(2p-8)-b\right\} \Phi(0,k_\perp) \Bigg]\, .
\een
Comparing this with \refb{ebrane2} for different values of $p$, we see that we must have:
\be
c=-1, \qquad a=0, \qquad b=2/\kappa\, .
\ee
We also see that if we had started with some arbitrary normalization on the right hand
sides of \refb{ebrane3} and \refb{ebrane3a}, the comparison between \refb{ebrane22} and
\refb{ebrane2} would have fixed them to be $\kappa \TT/2$ as given in 
\refb{ebrane3} and \refb{ebrane3a}, This in turn confirms the normalization of
\refb{e148}.

\end{document}